\documentclass[preprint,aps,singlecolumn,superscriptaddress,preprintnumbers,nopacs,overcite,floatfix,amsmath,amssymb]{revtex4}

\usepackage{graphicx}% Include figure files
\usepackage{dcolumn}% Align table columns on decimal point
\usepackage{bm}% bold math

\usepackage{amsmath}
\usepackage{graphics}
\usepackage{amssymb}
\usepackage{mathrsfs}
\usepackage{bbm}
\usepackage{epsfig}
\usepackage{makecell}
\usepackage{subfig}

\usepackage{array}
\usepackage{multirow}
\usepackage[justification=RaggedRight,singlelinecheck=false,font=small]{caption}
\usepackage{subfig}

\newcommand{\be}{\begin{eqnarray}}
\newcommand{\ee}{\end{eqnarray}}

\begin{document}
\vspace{1.5cm}

\title{\Large Virtual reality analysis of intrinsic protein geometry \\ with applications to {\it cis} 
peptide planes \\ }

\author{Yanzhen Hou}
\email{843528854@qq.com}
\affiliation{School of Physics, Beijing Institute of Technology, Beijing 100081, P.R. China}
\author{Jin Dai}
\email{daijing491@gmail.com}
\affiliation{School of Physics, Beijing Institute of Technology, Beijing 100081, P.R. China}
\author{Nevena Ilieva}
\email{nilieval@mail.cern.ch}
\affiliation{Institute of Information and Communication Technologies, Bulgarian Academy of Sciences, 25A, Acad. G. Bonchev Str., Sofia 1113, Bulgaria}
\author{Antti J. Niemi}
\email{Antti.Niemi@physics.uu.se}
\affiliation{School of Physics, Beijing Institute of Technology, Beijing 100081, P.R. China}
\affiliation{Nordita, Stockholm University, Roslagstullsbacken 23, SE-106 91 Stockholm, Sweden}
\affiliation{Department of Physics and Astronomy, Uppsala University,
P.O. Box 803, S-75108, Uppsala, Sweden}
\affiliation{Laboratoire de Mathematiques et Physique Theorique
CNRS UMR 6083, F\'ed\'eration Denis Poisson, Universit\'e de Tours,
Parc de Grandmont, F37200, Tours, France}
\affiliation{Laboratory of Physics of Living Matter, School of Biomedicine, Far Eastern Federal University, Vladivostok, Russia}
\homepage{http://www.folding-protein.org}
\author{Xubiao Peng}
\email{xubiaopeng@gmail.com}
\affiliation{Department of Physics and Astronomy, University of British Columbia, \\
Vancouver, British Columbia V6T1Z4, Canada}
\author{Jianfeng He}
\email{hjf@bit.edu.cn}
\affiliation{School of Physics, Beijing Institute of Technology, Beijing 100081, P.R. China}

\begin{abstract}
A protein is traditionally visualised as a piecewise linear discrete  
curve,  and its geometry is conventionally 
characterised by the extrinsically determined  Ramachandran angles. However, 
a protein backbone has also two independent intrinsic geometric structures, due to the 
peptide planes and the side chains. 
Here we adapt and develop modern 3D virtual reality techniques to scrutinize the atomic 
geometry along a protein backbone, in the vicinity of a peptide plane. 
For this we compare backbone geometry-based (extrinsic) and structure-based (intrinsic) coordinate systems, and
as an example we inspect  
the {\it trans} and {\it cis} peptide planes.
We reveal systematics in the way how a {\it cis} peptide plane deforms the neighbouring atomic geometry,
and we develop  a virtual reality based 
visual methodology that can identify the presence of a {\it cis} peptide plane from the arrangement of atoms in its vicinity. 
Our approach can easily detect exceptionally placed atoms in crystallographic structures. 
Thus it can be employed as a 
powerful visual refinement tool which is applicable 
also in the case when resolution of the protein structure is limited and whenever refinement is needed.
As concrete examples we identify a number 
of crystallographic protein structures in Protein Data Bank (PDB) 
that display exceptional atomic positions around their {\it cis} peptide planes. 
\end{abstract}

%\pacs{
%87.15.Cc 05.45.Yv 36.20.Ey
%}

\maketitle

\section{Introduction}
 
The visualisation of a three dimensional discrete framed curve is central to computer 
graphics. The issues range from the association of ribbons and tubes to the determination of camera 
gaze directions along trajectories. Common applications include various aspects of  aircraft and robot kinematics, stereo 
reconstruction and virtual reality \cite{Hanson-2006,Kuipers-1999}.
Here we propose to extend these modern  3D virtual reality techniques to analyse crystallographic protein structures.

Geometrically, a protein backbone is often visualised as a 
one-dimensional piecewise linear discrete chain, with vertices that customarily 
coincide with the positions of the C$^\alpha$ 
atoms. Indeed, for  a structureless one-dimensional 
chain the extrinsic and intrinsic geometries coincide.
However, in the case of a protein there is an essential supplement: A 
protein backbone has {\it a priori}  two {\it independent} intrinsic structures. One of these is specified by
the side chains  and the other relates to the peptide planes. Thus, in order to reliably 
determine the atomic anatomy of a protein we need to characterise not only 
the C$^\alpha$ backbone,
but we also need to know the side chain C$^\beta$ atoms and the peptide plane O atoms. At least within 
the assumption that the positions of the other heavy atoms can be reliably estimated using {\it e.g.} 
secondary structure or backbone dependent rotamer libraries \cite{Lovell-2000,Schrauber-1993,Dunbrack-1993,Shapovalov-2011,Peng-2014}.
Accordingly, in the case of  a protein, a combination of both extrinsic and intrinsic coordinate 
systems is needed, to provide a detailed visual insight to its structure.

Here we address the intrinsic geometry which is due to peptide planes; the intrinsic geometry that relates to C$^\beta$
and other side chain atoms has been analysed in [\citenum{Peng-2014}].
 Our results
demonstrate the value to go beyond the traditional approaches that visualise a protein structure 
either in terms of the (extrinsic) $ ...-\mathrm N-\mathrm C^\alpha - \mathrm C - \mathrm N - ...$ 
backbone based Ramachandran angles, 
or in terms of an (extrinsic) C$^\alpha$ backbone 
based laboratory  frame which is employed by most of the available 
3D protein visualisation programs including VMD, Jmol, PyMOL and many others \cite{wiki-viewer}. 
We first show how to utilise the peptide planes along a protein backbone to determine novel, intrinsic 
coordinates. We  then demonstrate how a detailed visual information on crystallographic protein  structure can be 
extracted in such a coordinate system. Our construction provides a basis for  the development of a future 
3D virtual reality based web-server,  to visually investigate the {\it intrinsic} assignment of atoms along a protein 
chain for the purpose of validation and
refinement, and in support of biomedical and pharmaceutical research.

%We are particularly interested in analysing the innate structure of  {\it cis}-$X--Xnp$ peptide planes.
As an application we consider the {\it cis} peptide planes in crystallographic protein structures:
In a protein, both the chemical composition and the geometric shape of a peptide plane are usually rigid, quite independently of the amino acid 
configuration \cite{Berkholz-2012,Hayward-2001}. A pair of  neighbouring C$^\alpha$ atoms defines two of the four `corners' of a peptide plane, 
and the other two coincide with the backbone O and H atoms; these four atoms are very tightly confined into a plane.
In the crystallographic structures in Protein Data Bank (PDB) \cite{Berman-2000}, 
the peptide plane is mostly  found in the {\it trans} conformation 
where the %(C$^{\alpha}_{i-1}$$-$C$_{i-1}$$-$N$_i$$-$C$^\alpha_{i}$)
Ramachandran dihedral $\omega$ between $\langle$C$^{\alpha}_{i-1}$$-$C$_{i-1}$$-$N$_i$$\rangle$ 
and $\langle$C$_{i-1}$$-$N$_i$$-$C$^\alpha_{i}$$\rangle$  planes has the value $\omega \approx \pi$. 
%(Note that we label all atoms in
%a given peptide plane with the same index.)  
The {\it cis} conformation where
$\omega \approx 0$ is also relatively common in the case of the imide bond X--Pro, but it is quite rare in the case of amide bonds X--Xnp, where X denotes any amino acid and Xnp is any residue except Pro \cite{Stewart-1990,Jabs-1999,Demange-2003,Berkholz-2012,Touw-2015}. It has been proposed
\cite{Jabs-1999} that
the relative rarity of  these  {\it cis} X--Xnp peptide planes in PDB structures 
could be partly due to the {\it a priori }
assumption which is commonly made during structure determination, that a peptide plane should be 
in a {\it trans} conformation. In fact,  the {\it cis} X--Xnp peptide planes are found to be 
significantly more common among high resolution structures, where the need for a refinement 
is smaller than in the case of the more refinement-prone medium and low resolution structures \cite{Stewart-1990,Jabs-1999}. 
Since the {\it cis} X--Xnp bond might be more abundant than previously thought, and since 
the majority of these  bonds seem to be located in the vicinity of functionally important regions \cite{Stoddard-1998},
there are good reasons to develop 
tools to visualise and analyse the way the  {\it cis} bonds affect the {\it intrinsic}  geometry of the neighbouring
atom structure in a protein.

%Here we aim to develop novel 3D visualisation tools to analyse the
%innate structure of  {\it cis}-$X--Xnp$ peptide planes.
%The   association of ribbons and tubes along a space curve and the determination of camera gaze directions
%along trajectories are both common tasks in various visualisation problems, from aircraft and robot kinematics to 
%stereo reconstruction and virtual reality \cite{Hanson-2006},  \cite{Kuipers-1999}. As a consequence
%the visualisation of a three dimensional discrete framed curve, akin the C$^\alpha$ backbone
%of a protein, is an important and widely studied topic in computer graphics. It is widely recognised 
%in the context of virtual reality, that there is an advantage to have a wide selection of 
%frames available when 3D-visualising a space curve \cite{Hanson-2006},  \cite{Kuipers-1999}. 
%Accordingly, our approach which combines extrinsically  and  intrinsically determined framings 
%along a protein chain,  
%is a complement both to the  traditional approach to visualise a protein structure in terms of Ramachandran angles, 
%and to modern 3D protein visualisation programs such as VMD, Jmol, PyMOL and others \cite{wiki-viewer} that commonly 
%employ an extrinsic  coordinate system which builds on the (extrinsic) laboratory   frame. 
%

\section{Methods}

%\subsection*{{\it cis}-peptide planes}
In the sequel we use the index $i=1,...,K$ to label a residue along a protein backbone with
$K$ residues, with indexing starting from the N terminus. 
We shall be interested in the $i^{th}$ 
peptide plane and its neighborhood. For a peptide plane that connects the residues 
$i$ and $i+1$, we shall use the same index $i$ to
label {\it all} the atoms, {\it i.e.} we use the indexing convention  
(... C$^{\alpha}_{i}$$-$C$_{i}$$-$N$_i$$-$C$^\alpha_{i+1}$ ... ). This   
deviates from the common convention,  to index an atom according to the ensuing amino acid in lieu of the adjacent
peptide plane.
We shall utilise
two different sets of coordinates. The first, {\it extrinsic} set of coordinates,
coincides with the C$^\alpha$ Frenet frames  introduced in \cite{Hu-2011}. 
%We use these coordinates to study, how the {\it cis}$X--Xnp$ 
%peptide planes affect the overall geometry of a protein backbone in $\mathbb R^3$. 
The second coordinate system --- the one we introduce here --- is {\it intrinsic}. We construct it 
using the C, N and O atoms of a given peptide plane: The C and N atoms appear as vertices along the 
piecewise linear discrete $ ...-\mathrm N-\mathrm C^\alpha - \mathrm C - \mathrm N - ...$ backbone, but 
the O atom does not: The C, N and O atoms then introduces an intrinsic framing of the backbone that 
we call the CNO frames.
%It yields us local information on the correlation 
%between side-chain and peptide plane geometry. 

%Note: In the sequel we shall identify all the atoms on the $i^{th}$ peptide plane with the same index $i$. This is
%a deviation from the convention to label the atoms according to the ensuing residue.

\subsection*{Discrete Frenet frames and Frenet spheres}
 
The discrete Frenet framing \cite{Hu-2011} is a purely geometric, extrinsic description of the protein backbone
that builds on the coordinates $\mathbf r_i$ of the C$^\alpha$ atoms. 
At  each vertex $\mathbf r_i$ of the backbone,
we define the discrete Frenet frame ($\mathbf t_i, \mathbf n_i , \mathbf b_i$) 
as follows: the $i^{th}$ tangent vector $\mathbf t_i$
 points from the center of the $i^{th}$ $\alpha$-carbon towards the center of the $(i+1)^{th}$  $\alpha$-carbon,
 \begin{equation}
 \mathbf t_i \ = \ \frac{\mathbf r_{i+1} - \mathbf r_i}{|\mathbf r_{i+1} - \mathbf r_i|} 
 \label{t}
 \end{equation}
 The binormal vector is
 \begin{equation}
 \hskip -0.45cm \mathbf b_i \ = \ \frac{ \mathbf t_{i-1} \times \mathbf t_i }{ |  \mathbf t_{i-1} \times \mathbf t_i |}
 \label{b}
 \end{equation}
 The normal vector is
 \begin{equation}
 \hskip -0.9cm \mathbf n_i \ = \ \mathbf b_i \times \mathbf t_i
 \label{n}
 \end{equation}
Together, these three vectors constitute a right-handed orthonormal frame, centered at the $i^{th}$ C$^\alpha$ atom.
In the sequel,  we shall 
use the convention that ($\mathbf n, \mathbf b,\mathbf t$) corresponds to the  
right-handed Cartesian ($xyz$)$\sim$($rgb$) coordinate system, with the convention that $x \sim \mathbf n$ is {\it 
green}
($g$), $y\sim \mathbf b$ is {\it blue} ($b$) and $z\sim \mathbf t$ is {\it red} $(r)$. 

\begin{figure}[h]
         \centering
                 \includegraphics[width=0.4\textwidth]{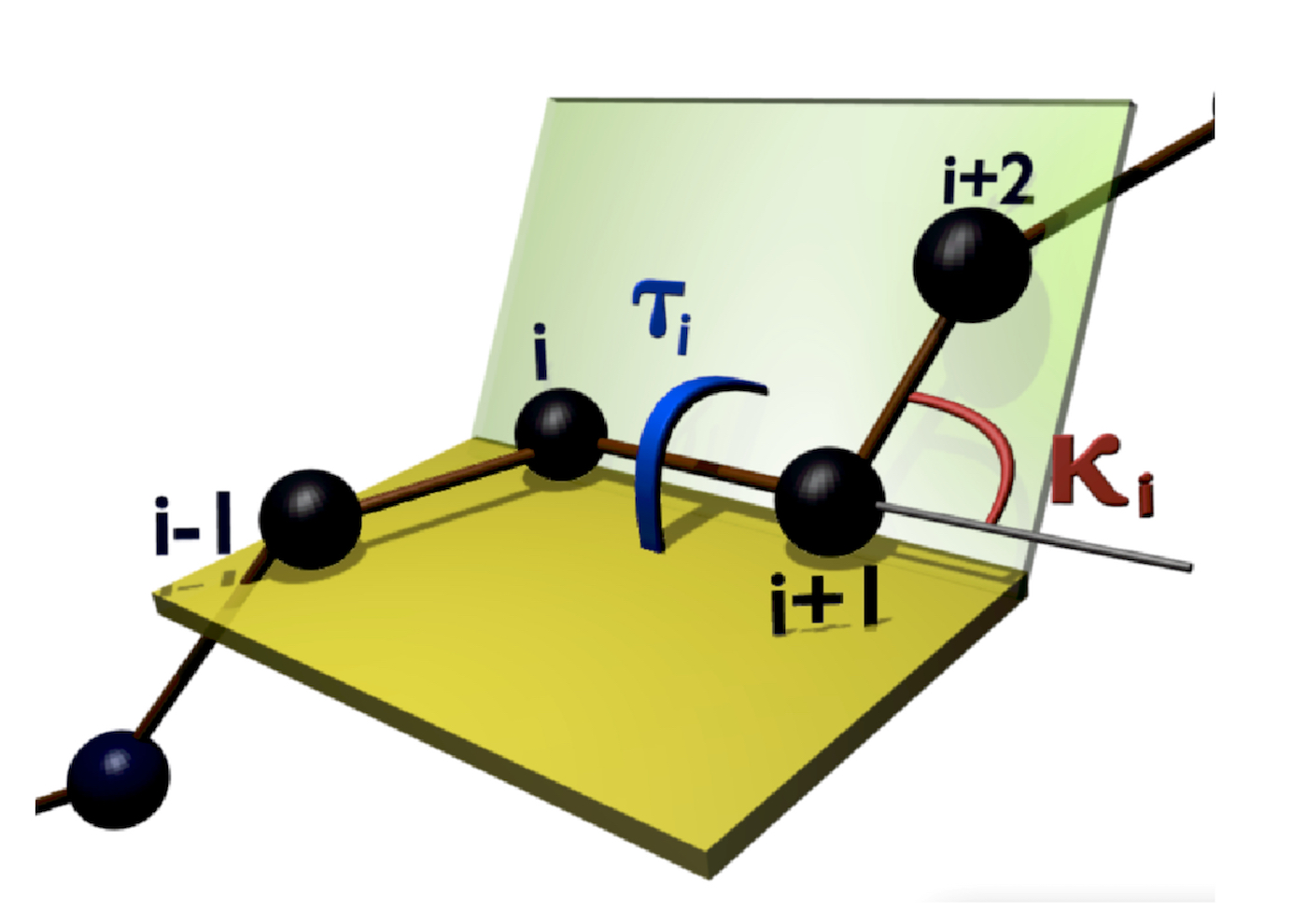}
         \caption{{ 
      (Color online) Bond ($\kappa_i$)  and torsion  ($\tau_i$)  angles (\ref{kappa}) and (\ref{tau}).
        }}
        \label{fig1}
 \end{figure}

We use the Frenet frames to define the virtual C$^\alpha$ backbone bond ($\kappa$) and torsion ($\tau$) angles as shown 
in Figure \ref{fig1},
%%%%%%%%%%%%%%%%%%%%%%%%%%%%%%%%%%%%%%%%%%%%%
%
%
%
%
%
%figure 13
%%%
%
%
%
%%%%%%%%%%%%%%%%%%%%%%%%%%%%%%%%%%%%%%%%%%%%%
%
%
\begin{equation}
\kappa_{i+1,i} \equiv \kappa_i \ = \  \arccos ( \mathbf t_{i+1} \cdot \mathbf t_i ) 
\label{kappa}
\end{equation}
\begin{equation}
\tau_{i+1,i} \equiv \tau_i \ = \ \sigma_i \arccos ( \mathbf b_{i+1} \cdot \mathbf b_i)
\label{tau}
\end{equation}
where 
\[
\sigma_i = {\rm sign}[ (\mathbf b_{i+1} \times \mathbf b_i) \cdot \mathbf t_i ]
\]
We identify the bond angle $\kappa \in [0,\pi]$ with the canonical 
latitude angle  of a (Frenet) two-sphere
which is centered at the C$^\alpha$ carbon; the tangent vector $\mathbf t$ points towards the
north-pole where $\kappa = 0$.  
The torsion angle $\tau\in [-\pi,\pi)$ is the longitudinal angle of the sphere, so that $\tau = 0$ 
on the great circle that passes both through the north pole and through the tip of the normal vector 
$\mathbf n \sim x$-axis, and the longitude increases in the counterclockwise direction 
around the  tangent vector. Accordingly ($\kappa,\tau$) are the standard spherical coordinates on the
Frenet two-sphere. 

\begin{figure}[h]
         \centering
                 \includegraphics[width=0.4\textwidth]{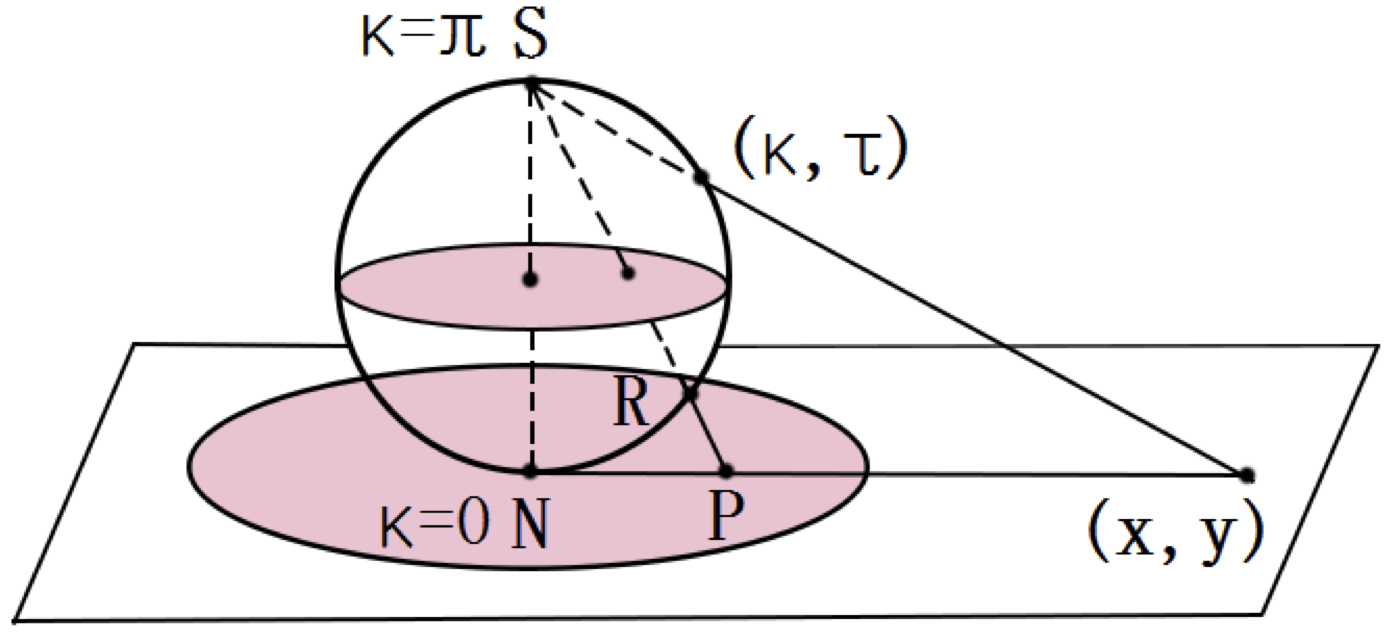}
         \caption{{ 
      (Color online) Stereographic projection of a two-sphere onto the Riemann sphere from the south pole.
        }}
        \label{fig2}
 \end{figure}

For visualisation, we find it occasionally convenient 
to stereographically project the sphere onto the complex ($x,y$) plane from the south-pole
\begin{equation}
x+iy \ \equiv \ \sqrt{ x^2 + y^2} \, e^{i\tau} \ = \ \tan\left( \kappa/2 \right) \, e^{i\tau}
\label{stereo}
\end{equation}
as shown in Figure \ref{fig2}. The north pole, where $\kappa=0$, becomes mapped to the origin ($x,y$)$=$($0,0$), while the south pole  
$\kappa=\pi$ is sent to infinity.

%
%%%%%%%%%%%%%%%%%%%%%%%%%%%%%%%%%%%%%%%%%%%%%
%
%
%
%
%
%figure   14
%%%
%
%
%
%%%%%%%%%%%%%%%%%%%%%%%%%%%%%%%%%%%%%%%%%%%%%
%
%

\subsection*{CNO frames and spheres}

A pair of  neighboring C$^\alpha$ atoms, located at sites $i$ and $i+1$,  defines  two corners of 
the $i^{th}$ peptide plane. The  two other corners 
coincide with the backbone O$_i$ and H$_i$ atoms. It has been found that in
crystallographic structures in Protein Data Bank (PDB), 
the various covalent bond lengths display very 
small variations from their average values, and the deviations of peptide planes from
a planar structure are very small. 
In particular, the N$_i$ and C$_i$ backbone atoms  lie tightly on the $i^{th}$ peptide plane. Moreover, the 
piecewise linear discrete $ ...-\mathrm N-\mathrm C^\alpha - \mathrm C - \mathrm N - ...$ chain
forms a covalently connected backbone of the protein, it determines a basis for introducing extrinsic coordinate systems 
such as the Ramachandran angles.

\begin{figure}[h]
         \centering
                 \includegraphics[width=0.4\textwidth]{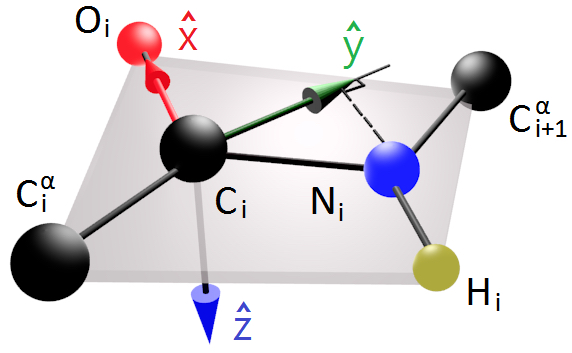}
         \caption{{ 
      (Color online) The right-handed orthonormal ($\mathbf x, \mathbf y, \mathbf z$)$\sim$ ($rgb$)
      CNO-frame is 
      defined in terms of the C, N and O atoms of the peptide plane. The Figure shows
      a {\it trans} X--Xnp conformation, the Ramachandran dihedral between the planes 
      $\langle$C$^{\alpha}_{i}$$-$C$_{i}$$-$N$_i$$\rangle$ and $\langle$C$_{i}$$-$N$_i$$-$C$^\alpha_{i+1}$$\rangle$ 
      has the value $\omega = \pi$.
        }}
        \label{fig3}
 \end{figure}

%The observed variations in the positioning of the C$_i$, N$_i$ and O$_i$ atoms 
%of the peptide plane are found to be  {\it very} 
%small,  their relative arrangement appears to be essentially independent of the adjacent 
%amino acids and of the protein structure. 
Here we shall  employ  the three  atoms C$_i$, N$_i$ and O$_i$ of a given peptide plane 
to define a framing along the  backbone chain. Since the O$_i$ atom is 
not located along the $ ...-\mathrm N_{i-1}-\mathrm C_i^\alpha - \mathrm C_i - \mathrm N_i - ...$  backbone, such a framing
is {\it intrinsic} to the protein. It determines a basis for introducing intrinsic, structure based 
coordinate systems. Accordingly, we introduce an orthonormal CNO-frame with 
($\mathbf x, \mathbf y, \mathbf z$) the ensuing right-handed Cartesian ($xyz$)$\sim$($rgb$) orthonormal basis; we
place the origin at the location of the C$_i$ atom on the peptide plane. Specifically, 
let $\mathbf r_{\mathrm C i}$, $\mathbf r_{\mathrm  N i}$ and $\mathbf r_{\mathrm  O i}$ be the coordinates of the
C$_i$, N$_i$ and O$_i$ atoms on the $i^{th}$ peptide plane. Then, we define the CNO frame as follows:
\begin{equation}
\mathbf x_i \ = \ \frac{ \mathbf r_{\mathrm O i} - \mathbf r_{\mathrm C i} }{ | \mathbf r_{\mathrm
O i} - \mathbf r_{\mathrm C i} |} \ \ , \ \ \ 
\mathbf z_i \ = \ \frac{ \mathbf x_i \times \mathbf u_i  }{ |  \mathbf x_i \times {\mathbf u}_i  |}
\ \ , \ \ \ \mathbf y_i \ = \ \frac{ \mathbf z_i \times \mathbf x_i  }{ |  \mathbf z_i \times \mathbf x_i  |}
\label{uvw}
\end{equation}
where
\[
{\mathbf u}_i \ = \ \frac{ \mathbf r_{\mathrm N i} - \mathbf r_{\mathrm C i} }{ | \mathbf r_{\mathrm
N i} - \mathbf r_{\mathrm C i} |}
\]
See Figure \ref{fig3}, where  in particular we employ the color assignment ($rgb$) for the Cartesian ($xyz$) axes.

%%%%%%%%%%%%%%%%%%%%%%%%%%%%%%%%%%%%%%%%%%%%%
%
%
%
%
%
%figure   15
%%%
%
%
%
%%%%%%%%%%%%%%%%%%%%%%%%%%%%%%%%%%%%%%%%%%%%%
%
%

In analogy with the Frenet sphere we may also introduce a unit radius 
CNO two-sphere, when need be. This sphere is centered at the C atom of the peptide plane and in the sequel 
we denote by 
($\theta, \varphi$) the  ensuing latitude and longitude angles.

\section{Results}

We have studied in detail a subset of crystallographic PDB proteins, that
%Nature Structural Biology 10, 980 (2003)
%doi: 10.1038/nsb1203-980
consists of all presently available ultra high resolution structures that have been  
obtained using diffraction data with better than 1.0 \AA~ resolution; there are 557 protein structures
in our data set. We have chosen this PDB subset with the presumption, that in the case of high resolution structures the need for refinement in determining the various atomic coordinates
is minimal.

\subsection*{The {\it cis} peptide planes}

Here we define a peptide plane to be  in the {\it cis}-position when the Ramachandran 
angle $\omega$ has a value between $-\pi/4 < \omega < \pi/4$. We have inspected the 
PDB structures to conclude that this is a reasonable criterion; see \cite{Berkholz-2012}.
We have identified a total of 383 such {\it cis}-peptide plane structures in our data set. This  proposes that around 70$\%$ of proteins
have at least one such peptide plane. Of these, 
329 involve a proline in the $(i+1)^{th}$  corner, i.e. have the structure {\it cis} X--Pro.
This leaves us with 54 {\it cis} X--Xnp peptide planes
for our final analysis.  Thus our statistics suggests that as many as around 10$\%$ of high resolution proteins in 
PDB might  support a {\it cis} X--Xnp peptide plane.
%apparently located  near a functionally important region.

%%%%%%%%%%%%%%%%%%%%%%%%%%%%%%%%%%%%%%%%%%%%%
%
%
%
%
%
%figure 1
\begin{figure}[h]         
\centering            
  \resizebox{8 cm}{!}{\includegraphics[]{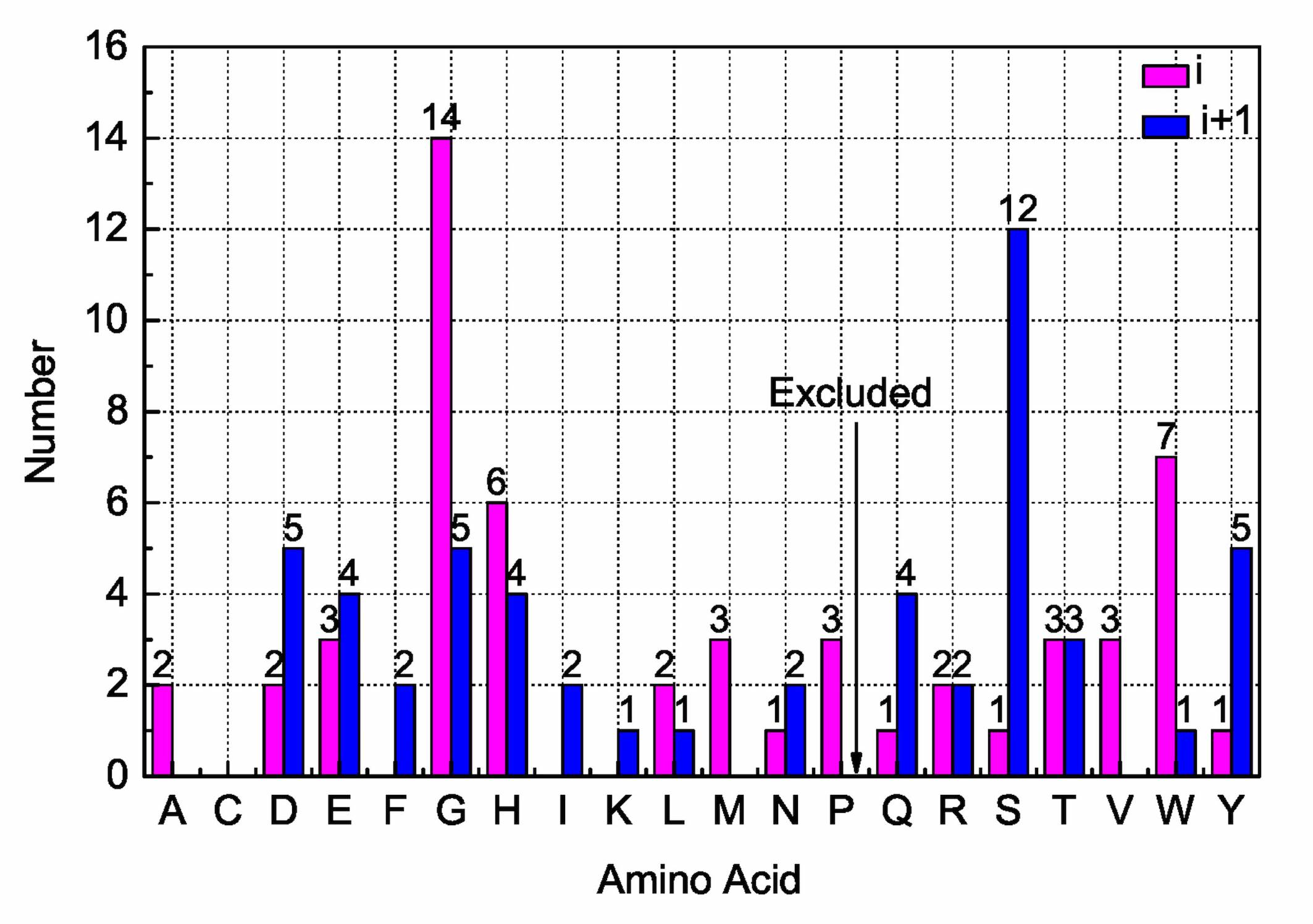}}
\caption {\small  (Color online) Distribution of amino acids at the $i^{th}$ and $(i+1)^{th}$ vertex of the {\it cis} X--Xnp peptide planes. Note that we consider only structures that exclude  proline from the $(i+1)^{th}$ vertex. }   
\label{fig4}    
\end{figure}
%%%
%
%

Figure \ref{fig4} 
shows the amino acid distribution of {\it cis} X--Xnp peptide planes in our data set, separately at the $i^{th}$ and
at the $(i+1)^{th}$ vertex; note that we do not exclude proline at the $i^{th}$ vertex, it appears three times in our data.
We find all amino acids except Cys in these peptide planes, but the statistics is too limited for us to make a conclusion on Cys, so
we look forward for an example.
It is notable that Gly has very high propensity to the  $i^{th}$
vertex,  while Ser appears
to be relatively common in the $(i+1)^{th}$ vertex. We also observe that 
unlike the case of the $(i+1)^{th}$ vertex where Pro is relatively abundant,
examples of $i^{th}$ vertex {\it cis}-Pro are quite rare.

%%%%%%%%%%%%%%%%%%%%%%%%%%%%%%%%%%%%%%%%%%%%%
%
%
%
%
%
%figure 2
%\begin{figure}
%\captionsetup[subfigure]{labelformat=empty}
%\centering
%\subfloat[]{\includegraphics[width=0.15\textwidth]{figure-2a.jpg}}
%\subfloat[]{\includegraphics[width=0.20\textwidth]{figure-2b.jpg}}
%\vskip -0.5cm
%\caption{(Color online) Distribution of Ramachandran $\omega$ values in our set of {\it cis} X--Xnp peptide planes, with  a Gaussian distribution and its one standard deviation distance denoted. We have labeled those entries that we shall
%         follow as examples; the corresponding PDB codes are identified in Table \ref{table-1}. In our statistical analysis we consider  structures within $-\pi/4 < \omega < \pi/4$, shown in dotted lines (left panel); we take the example F from outside of this range.
%         Distribution of C$^\alpha$--C$^\alpha$ distances in our set of {\it cis} X--Xnp peptide planes  are also shown (right panel).}
%    \label{fig2}
%  \end{figure}
%
%
%
%
\begin{figure}[h]
         \centering
        % \subfigure
           {\includegraphics[width=0.4\textwidth]{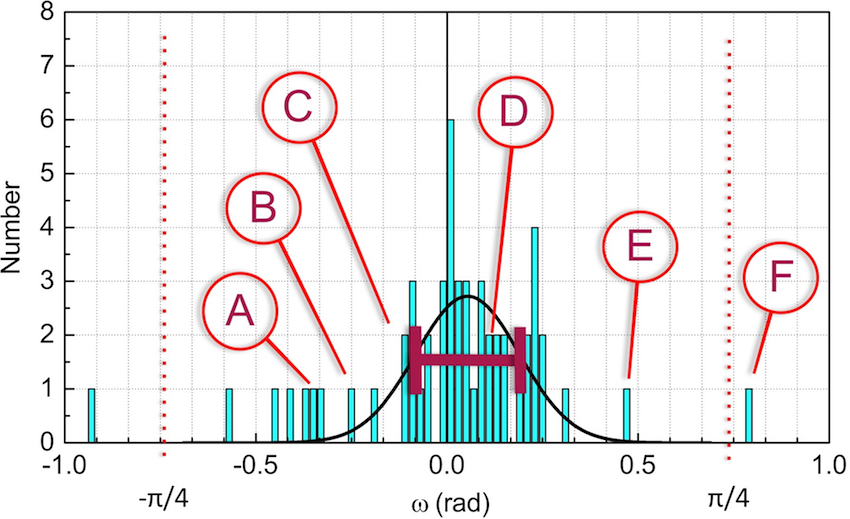} }
           \qquad
         % \subfigure
           {\includegraphics[width=0.4\textwidth]{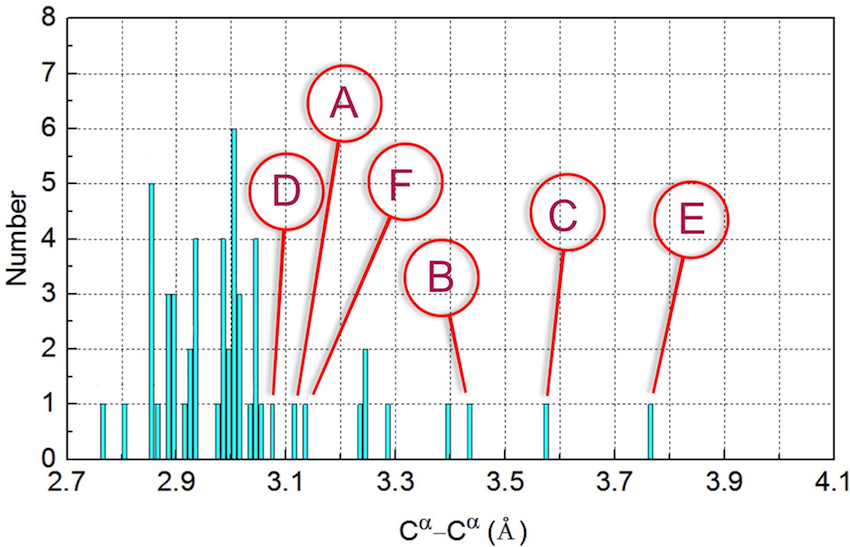} }
         \caption{{ (Color online) Distribution of Ramachandran $\omega$ values in our set of {\it cis} X--Xnp peptide planes, with  a Gaussian distribution and its one standard deviation distance denoted. We have labeled those entries that we shall
         follow as examples; the corresponding PDB codes are identified in Table \ref{table-1}. In our statistical analysis we consider  structures within $-\pi/4 < \omega < \pi/4$, shown in dotted lines (left panel); we take the example F from outside of this range.
         Distribution of C$^\alpha$--C$^\alpha$ distances in our set of {\it cis} X--Xnp peptide planes  are also shown (right panel).}}
       \label{fig5}
\end{figure}
%%
%
%
%
%%%%%%%%%%%%%%%%%%%%%%%%%%%%%%%%%%%%%%%%%%%%%
%
%
In Figure \ref{fig5} (left panel) we show the distribution of the Ramachandran dihedral $\omega$ for values
$-1<\omega<1$ in our data set. 
The corresponding bond lengths are shown in the right panel.  
We observe that there is a gap in values of $\omega$ around $\omega = \pm \pi/4$.
This justifies our choice of definition of {\it cis}-structures, for the present study:
Only structures with $|\omega| < \pi/4$ are included in our statistics.
See  however, the analysis in \cite{Berkholz-2012}.

In Figures \ref{fig5}  we have also identified a number of entries, these are the examples that we shall follow in the sequel, to 
exemplify how the approach can be used to detect and identify atoms with 
apparently exceptional positions; we do not analyse the (biological) consequences. 
We note that each of the example we have chosen is in some sense exceptional: Either the value of the
Ramachandran dihedral deviates from $\omega = 0$ by more than one standard deviation, or as in the case of 
2UU8$\_$A$\_$120-121 and 1NLS$\_$A$\_$121-122 (example B and D),  the preceding peptide plane is also {\it cis} X--Xnp.
The  PDB codes of our examples are listed in Table \ref{table-1}.

\begin{table}[h]
\renewcommand\arraystretch{1.5}
\centering
\begin{tabular}{clc}
\hline
Label &                           ~~~    PDB Entry \\                                       
\hline
\hline
A&                                                      2UU8$\_$A$\_$121-122 \\                                              
\hline
B&                                                      2UU8$\_$A$\_$120-121 \\                                               
\hline
C&                                                    1NLS$\_$A$\_$120-121 \\                                              
\hline
D&                                                   1NLS$\_$A$\_$121-122 \\                                               
\hline
E&                                                  1LUG$\_$A$\_$3-4 \\                                            
\hline
F&                                                   2GGC$\_$A$\_$191-192 \\                                        
\hline
\end{tabular}
\caption{The PDB entries that we follow as examples; first residue appears in $i^{th}$, the second in $(i+1)^{th}$ vertex. Note that 
2UU8 and 1NLS have two consecutive {\it cis} X--Xnp peptide planes, and 1LUG$\_$3 is glycine.}
\label{table-1}
\end{table}
%
%
%
%
%
%
%

%\vspace{0.3cm}

\subsection*{(Extrinsic) Frenet frames and the {\rm C}$^\alpha$ backbone}

\begin{figure}[h]
        \centering
        % \subfigure
          {\includegraphics[width=0.4\textwidth]{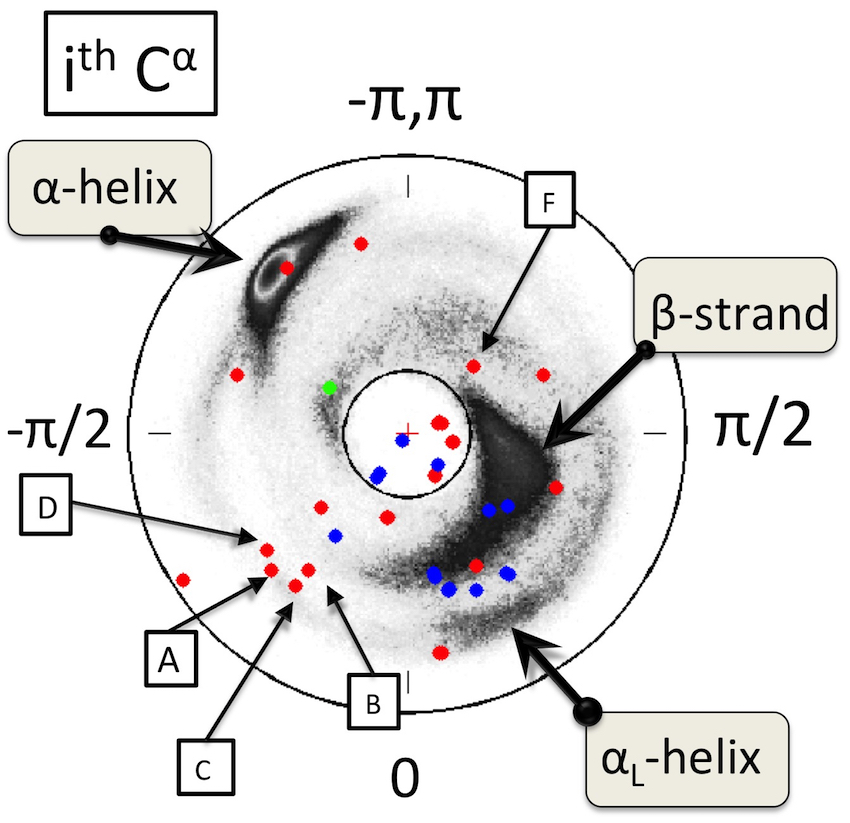} }
        % \subfigure
          {\includegraphics[width=0.4\textwidth]{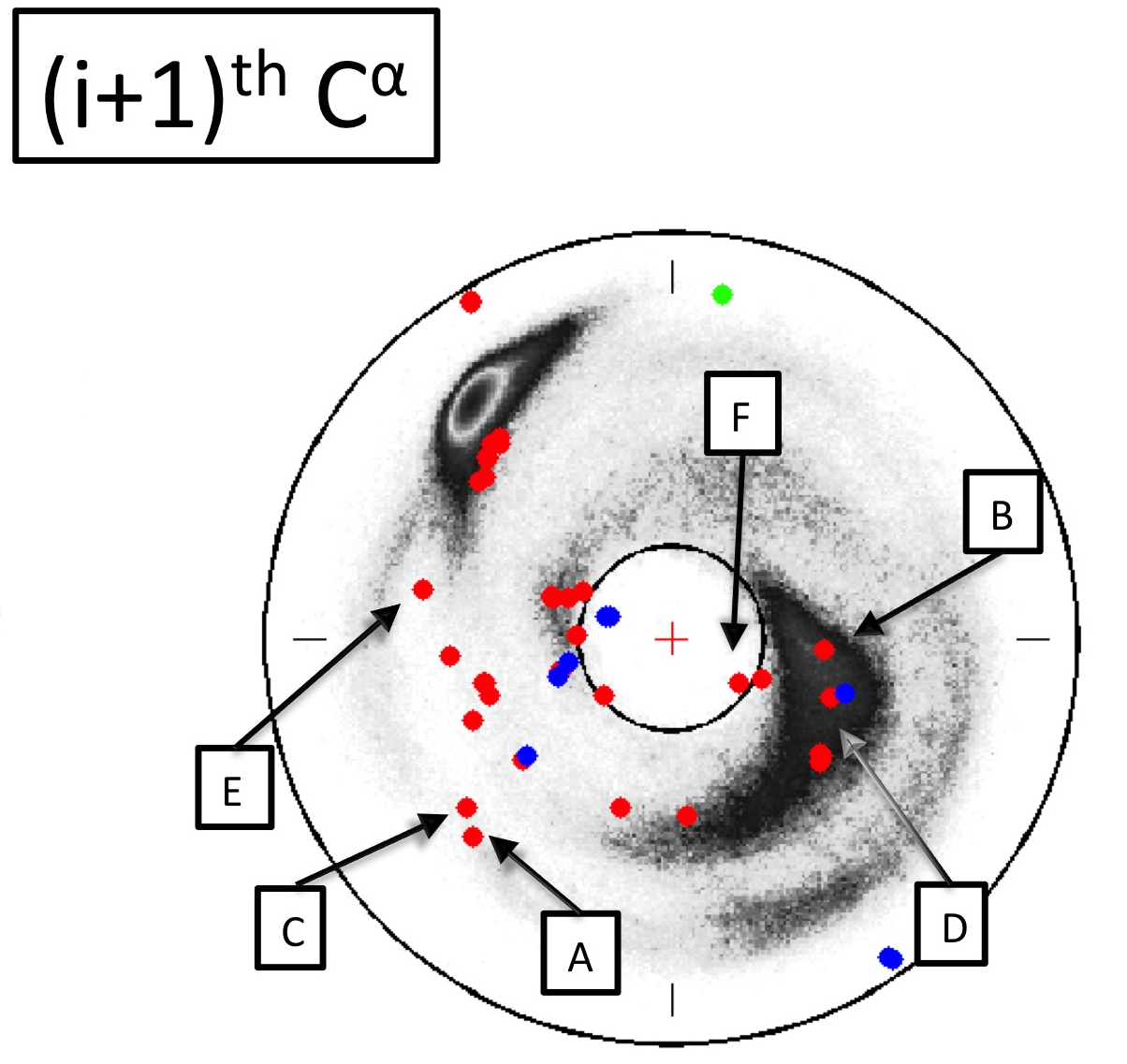} }
        \caption{{(Color online) Distribution of {\it cis} X--Xnp in the C$^\alpha$ Frenet frames, on the stereographic projection \ref{fig2}. 
        Major secondary structure regions ($\alpha$-helix, $\beta$-strand, left-handed $\alpha$-helix) are identified.
        Green is classified as helix, blue is strand and red is loop according to \cite{uniprot}. The two panels show ($\kappa,\tau$) at $i^{th}$ (left)  resp.   $(i+1)^{th}$ (right) vertex of the C$^\alpha$ backbone, as defined by (\ref{kappa}), (\ref{tau}). }}
  \label{fig6}
  \end{figure}

%We refer to the Methods Section, for the concepts that we employ in the sequel: 
The Figures \ref{fig6} show the distribution of the bond $\kappa_i$ and torsion $\tau_i$ angles of the 
{\it cis} X--Xnp entries in our statistical set, on the C$^\alpha$ centered Frenet spheres; the angles are defined in  
equations (\ref{kappa}),  (\ref{tau})  and Figure \ref{fig1}. We use in Figures \ref{fig6} 
the stereographic projection (\ref{stereo}) (Figure \ref{fig2}). The notion of  Frenet sphere and its generalisations in the context of 
protein visualisation, have been introduced and analysed in \cite{Hu-2011,Lundgren-2012,Lundgren-2012b}.
The center of the annulus in Figures \ref{fig6} corresponds to $\kappa=0$, while
points at infinite distance on the plane have $\kappa=\pi$. The torsion angle $\tau$ concurs with rotation around the center
of the annulus, its  values are marked in the left Figure \ref{fig6}. The grey background is the ($\kappa_i,\tau_i$) distribution 
for all PDB structures with resolution below 1.0 \AA. The majority of the structures are located in an annulus,
between two circles. The outer boundary circle corresponds to  $\kappa \sim \pi/2$. Structures
with larger values of $\kappa$ are sterically excluded, due to a clash between the neighboring C$^\alpha$ atoms. 
The inner boundary circle denotes 
$\kappa \sim 1$. Structures with a smaller value of $\kappa$ are sterically allowed but rare in PDB. The colored dots denote the individual ($\kappa_i,\tau_i$)  values of the
{\it cis} X--Xnp peptide planes. The  coloring is assigned 
according to the classification we obtain from
Uniprot \cite{uniprot}:  Green is classified as helix, 
blue is classified as strand, and red is coil according to \cite{uniprot}. The Figure \ref{fig6}  left
denotes the ($\kappa,\tau$) values that are assigned to the $i^{th}$ vertex, 
and the one on right denotes the ($\kappa,\tau$) values that are 
assigned to the $(i+1)^{th}$ vertex of the peptide plane. 

%%%%%%%%%%%%%%%%%%%%%%%%%%%%%%%%%%%%%%%%%%%%%
%
%
%
%
%
%figure 3

%%%%%%%%%%%%%%%%%%%%%%%%%%%%%%%%%%%%%%%%%%%%%
%
%

We observe  that the  
($\kappa,\tau$) values of {\it cis} X--Xnp peptide planes are common in the
region between $\tau \in [0,-\pi/2]$ which is otherwise rare in the PDB structures. Moreover, there is a clear
propensity towards $\kappa < 1$ in particular at the 
$i^{th}$ C$^\alpha$ vertex while these bond angle values are quite rare among {\it trans} peptide planes. We note that
according to (\ref{t}), (\ref{kappa}) a small value of $\kappa$ implies geometrically that there  is a 
tendency for the C$^\alpha$ backbone to straighten itself.
%at the {\it cis}$X--Xnp$ peptide plane. 
%Thus the {\it cis}$X--Xnp$ peptide planes are relatively often associated with abnormalities in the C$^\alpha$
%backbone geometry. 
We observe certain mismatch between the regular secondary structure assignment  
of a residue  according to \cite{uniprot}, and the actual placement of the dots in Figures \ref{fig6}.
%In Figure \ref{fig3}, we have identified some  {\it cis} X--Xnp 
%PDB residues (PDB code, chain, residue number) as representative examples.
%We shall keep following these examples in the sequel.

%%%%%%%%%%%%%%%%%%%%%%%%%%%%%%%%%%%%%%%%%%%%%
\vspace{0.2cm}

\subsection*{Side chains and peptide planes in Frenet frames}

\begin{figure}[h]
         \centering   
          % \subfigure
           {\includegraphics[width=0.4\textwidth]{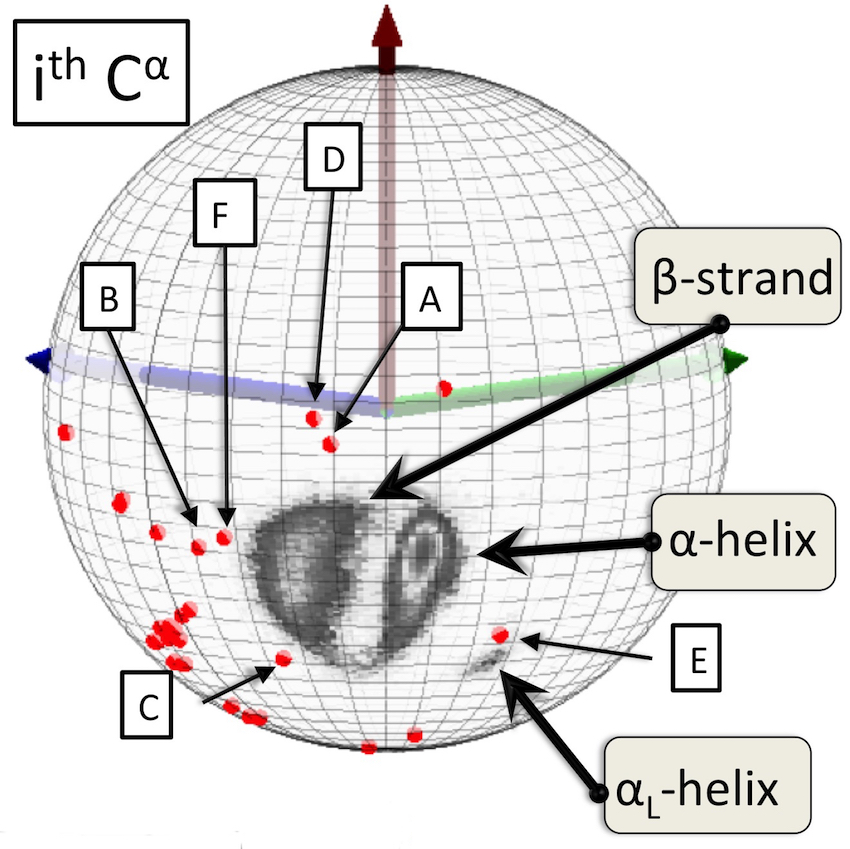}}
           \quad\quad\quad
          % \subfigure
           {\includegraphics[width=0.4\textwidth]{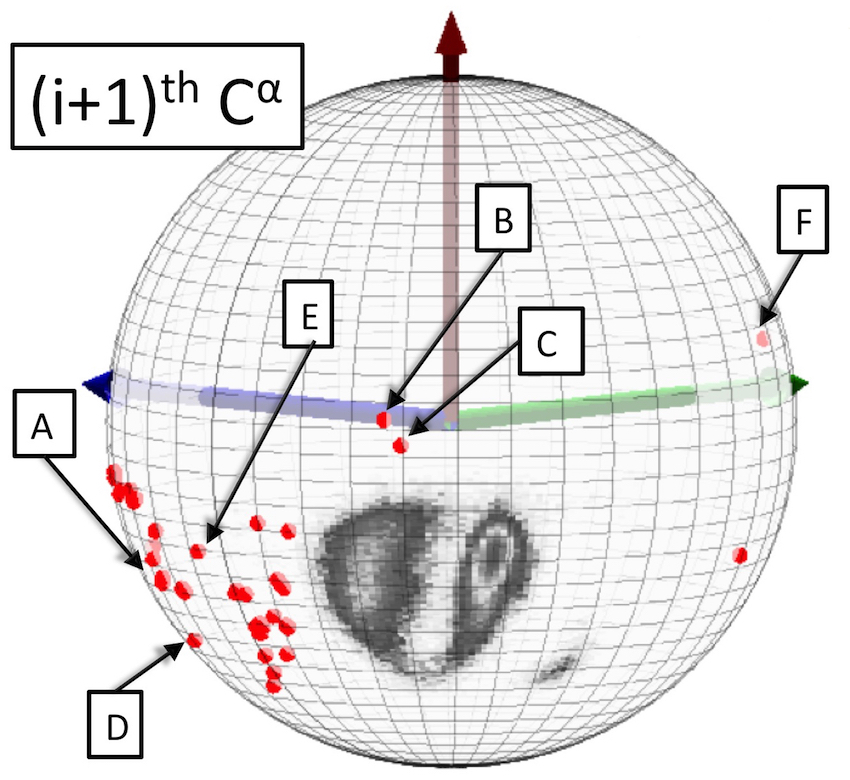}}
         \caption{(Color online) The distribution of side-chain C$^\beta$ in terms of the  ($\kappa,\tau$) angles on the Frenet sphere, for {\it cis} X--Xnp residues, for the $i^{th}$ C$^\beta$ (left) and the $(i+1)^{th}$  C$^\beta$ (right).}
        \label{fig7}
 \end{figure}

\begin{figure}[h] %{\textwidth} %[h]
         \centering
         % \subfigure
        {\includegraphics[width=0.38\textwidth]{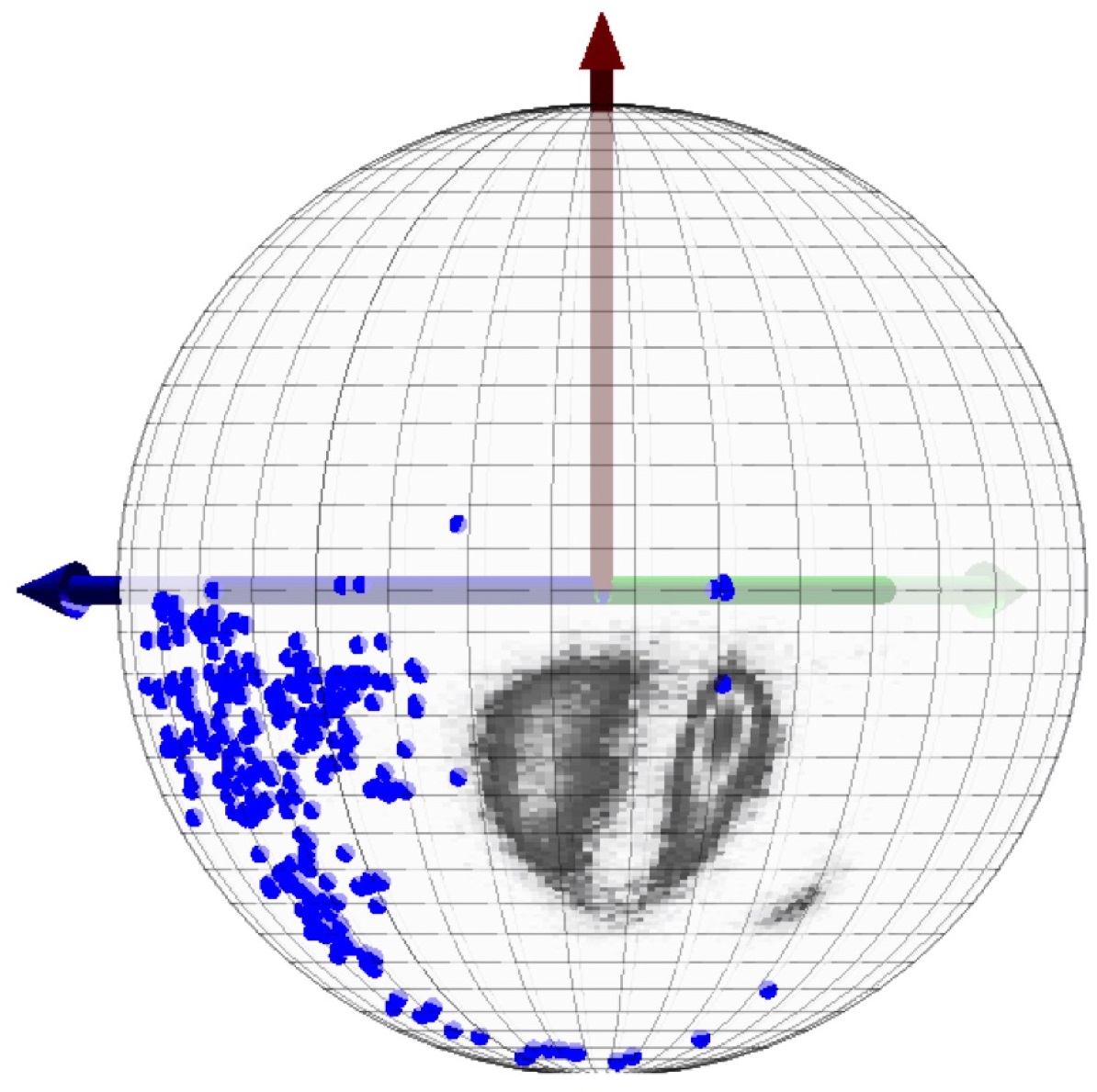}}
        \qquad\qquad\quad
          % \subfigure
          {\includegraphics[width=0.38\textwidth]{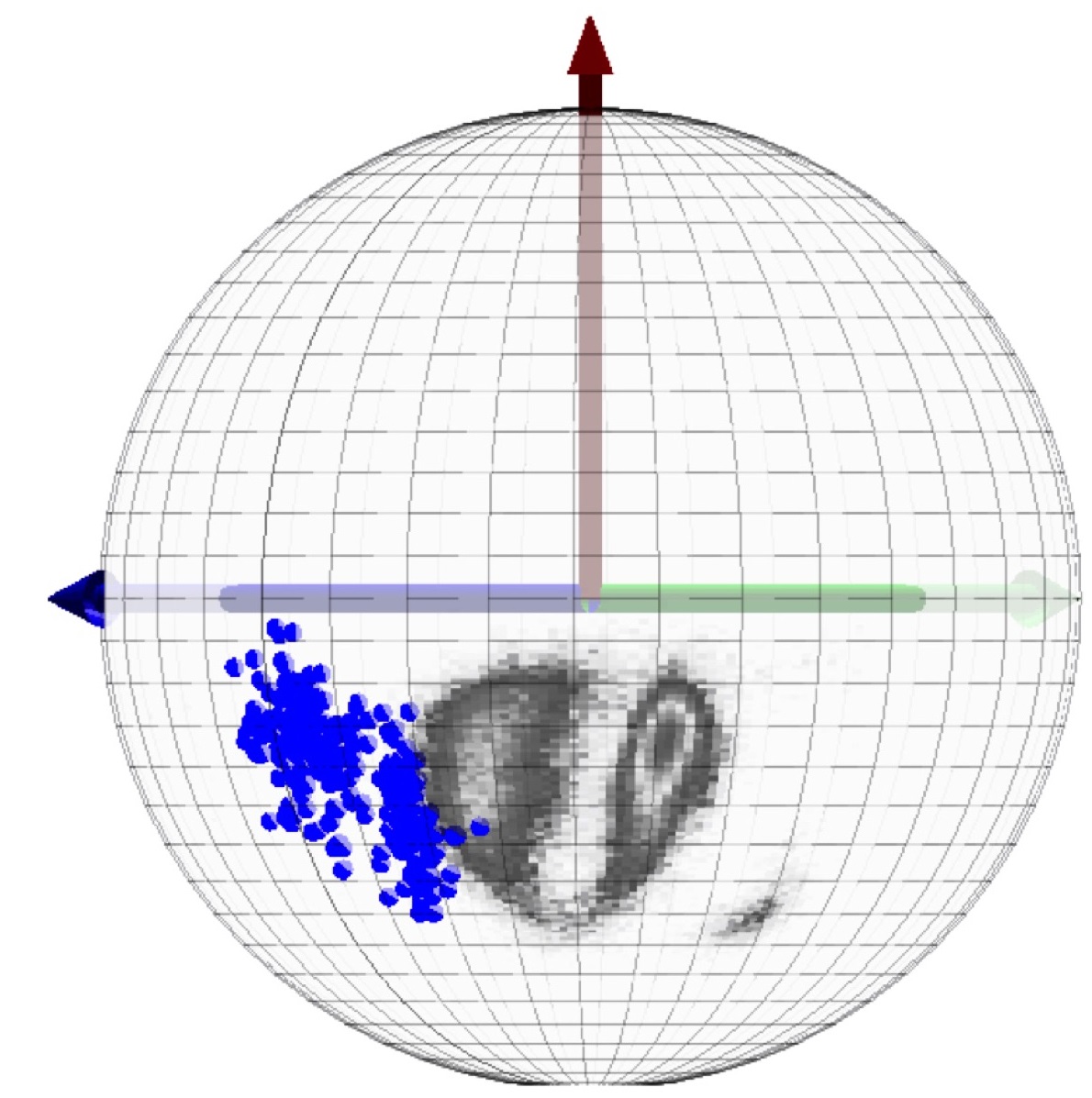}} 
         \caption{(Color online) Same as in Figure \ref{fig7} but for {\it cis} Pro--X;  (left) for the $i^{th}$ C$^\beta$ and (right)
         for the $(i+1)^{th}$  C$^\beta$.} 
        \label{fig8}
 \end{figure}

\begin{figure}[h]
          \centering
          % \subfigure
          {\includegraphics[width=0.4\textwidth]{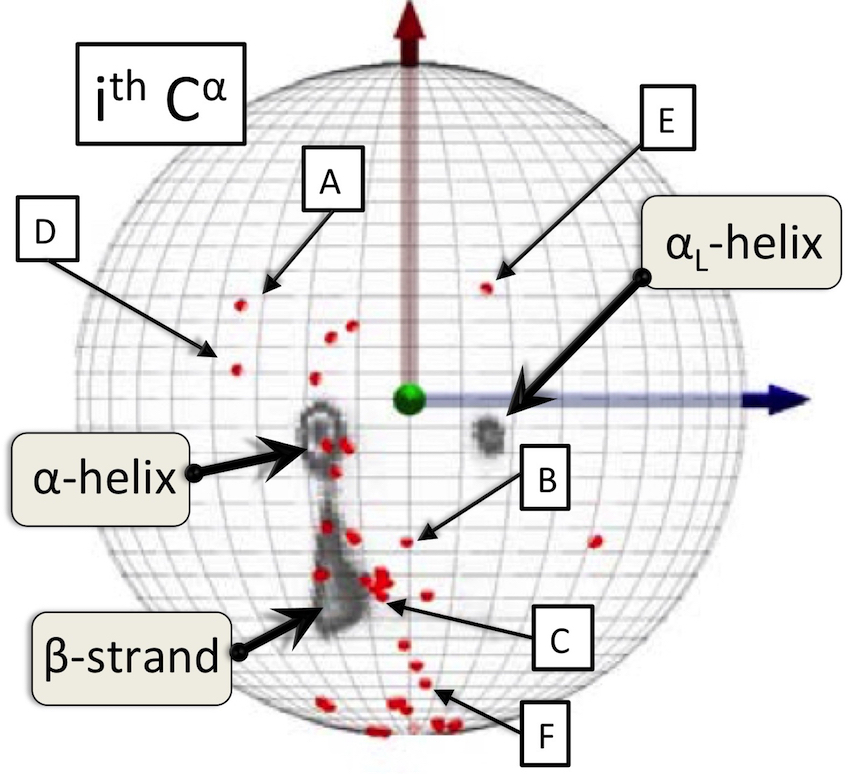}}
          \quad\quad\quad
           % \subfigure
          {\includegraphics[width=0.4\textwidth]{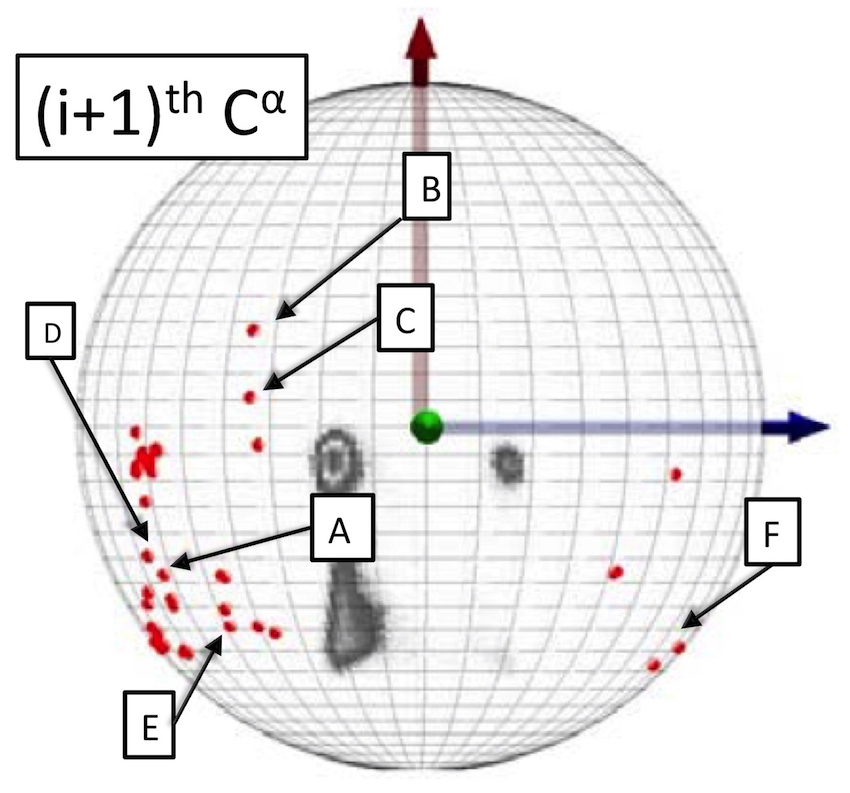}}
          \caption{(Color online) Same as in Figure \ref{fig7}, but for the N atom covalently bonded to C$^\alpha$, as seen from the $i^{th}$ (left panel) and $(i+1)^{th}$ (right panel) C$^\alpha$ atom, respectively.}%
         \label{fig9}
  \end{figure}

We proceed to visually investigate 
the individual C$^\beta$, N and  C  atoms that are covalently bonded to a C$^\alpha$ atom at either the 
$i^{th}$ or the $(i+1)^{th}$ vertex of the peptide plane. The results for the {\it cis} X - Xnp  peptide planes
in our statistical background are displayed in Figures \ref{fig7}, \ref{fig8} and \ref{fig9};   
the grey background is the ($\kappa,\tau$) distribution 
for all PDB structures with resolution below 1.0 \AA. 
In each of these figures, % Figure \ref{fig7}, \ref{fig9} and \ref{fig10} 
we have denoted the six residues as examples, that 
we have previously identified  in Figures \ref{fig5}, \ref{fig6} and listed in Table \ref{table-1}. 

%
%
%%%%%%%%%%%%%%%%%%%%%%%%%%%%%%%%%%%%%%%%%%%%%
%
%
%
%
%
%figure   4
%%%
%
%
%
%%%%%%%%%%%%%%%%%%%%%%%%%%%%%%%%%%%%%%%%%%%%%
%
%

Figure \ref{fig7}  shows the distribution of the $i^{th}$ and $(i+1)^{th}$ C$^\beta$ atoms, on the C$^\alpha$ centered
Frenet sphere.  It is notable how the distributions are consistently shifted towards left, in the Figure. 
For comparison, in Figure \ref{fig8} we show the same distributions in the case of  {\it cis} Pro--X. 
%%%%%%%%%%%%%%%%%%%%%%%%%%%%%%%%%%%%%%%%%%%%%
%
%
%
%
%
%figure  5
%%%
%
%
%
%%%%%%%%%%%%%%%%%%%%%%%%%%%%%%%%%%%%%%%%%%%%%
%
%
We observe the same overall pattern of shift, towards left in the Figure. We also observe that
in the case of $i^{th}$ C$^\beta$ the dispersion in both {\it cis} X--Xnp and {\it cis} Pro--X distributions is higher, than in the case of the $(i+1)^{th}$ C$^\beta$. Thus, the $i^{th}$ vertex 
 of a {\it cis} peptide plane seems to be
 more strained. 
 
 %%%%%%%%%%%%%%%%%%%%%%%%%%%%%%%%%%%%%%%%%%%%%
 %
 %
 %
 %
 %
 %figure 6
 %%%
 %
 %
 %
 %%%%%%%%%%%%%%%%%%%%%%%%%%%%%%%%%%%%%%%%%%%%%
 %
 %

\begin{figure}[h]
         \centering
         \includegraphics[width=0.4\textwidth]{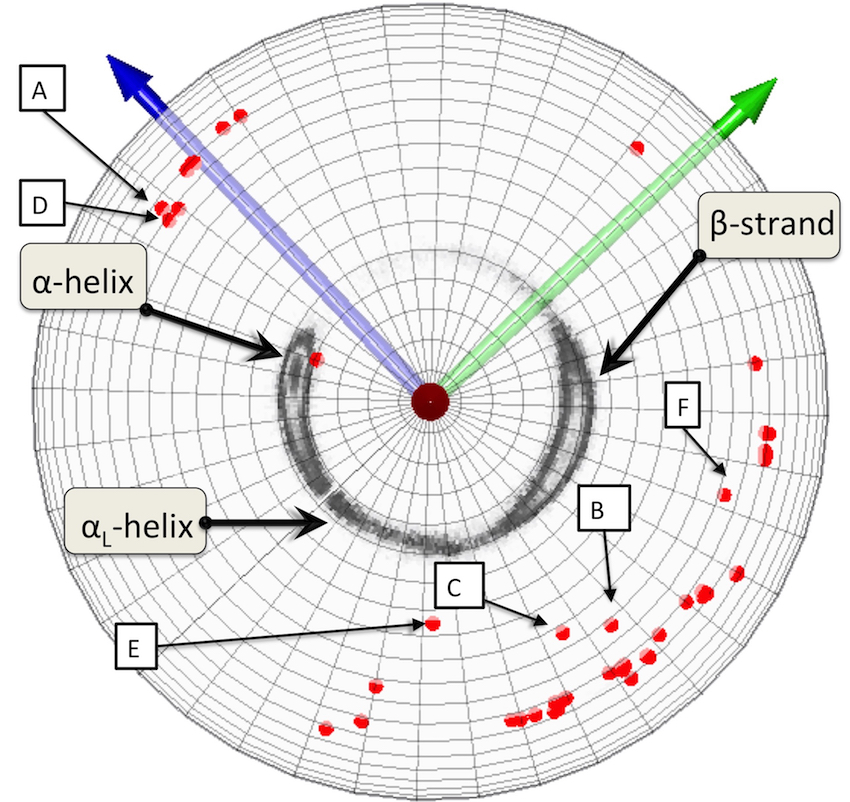}%
         \caption{(Color online) Same as in Figure \ref{fig7} (top) but for the C atom covalently bonded to the $i^{th}$ C$^\alpha$. In the $(i+1)^{th}$ Frenet frame we do not observe much deviation from the background, thus we do not display it. } %
        \label{fig10}
 \end{figure}

In Figure \ref{fig9} we show the $i^{th}$ and $(i+1)^{th}$ N atoms, on the C$^\alpha$ centered Frenet sphere.  
We observe that there is a quite strong dispersion in both distributions, 
the overall pattern is clearly different than the background. 
Finally, in Figure \ref{fig10} we have the $i^{th}$ C atoms, on the C$^\alpha$ centered
Frenet sphere. The distribution of the $(i+1)^{th}$ C atoms does not show much deviation 
from the background, thus is not displayed. Whether this is due to refinement practices or actual observations,  can not be determined
on the basis of the existing PDB data. 
%
%
%
%
%
%%%%%%%%%%%%%%%%%%%%%%%%%%%%%%%%%%%%%%%%%%%%%
%
%
%
%
%
%figure 7
%%%
%
%
%
%%%%%%%%%%%%%%%%%%%%%%%%%%%%%%%%%%%%%%%%%%%%%
%
%

\subsection*{(Intrinsic) CNO frames}

The CNO frames are peptide plane specific, and as such {\it intrinsic} 
to the protein structure; as shown in Figure \ref{fig3} 
the CNO coordinate system 
engages the C, N and O atoms of a single peptide plane.
Accordingly we  expect the CNO frames to provide us with intrinsic structural information, in particular
on the residues that are adjacent to the peptide planes. As examples, we consider
the C$_i^\alpha$ and C$_{i+1}^\alpha$ atoms at the two vertices together with  
the ensuing covalently  bonded atoms C$_i^\beta$, C$_{i+1}^\beta$, N$_{i-1}$ and  C$_{i+1}$, the
two latter engage the neighboring peptide planes. We note that the N$_{i-1}$, C$_i^\alpha$, C$_{i+1}^\alpha$ and C$_{i+1}$
are all engaged in the definition of the Ramachandran angles, that relate to the $i^{th}$ peptide plane. Thus no information can be gained of these atoms, in terms of Ramachandran angles.

%%%%%%%%%%%%%%%%%%%%%%%%%%%%%%%%%%%%%%%%%%%%%
%
%
%
%
%
%figure 8
 \begin{figure}[h]
         \centering 
         % \subfigure
         {\includegraphics[width=0.4\textwidth]{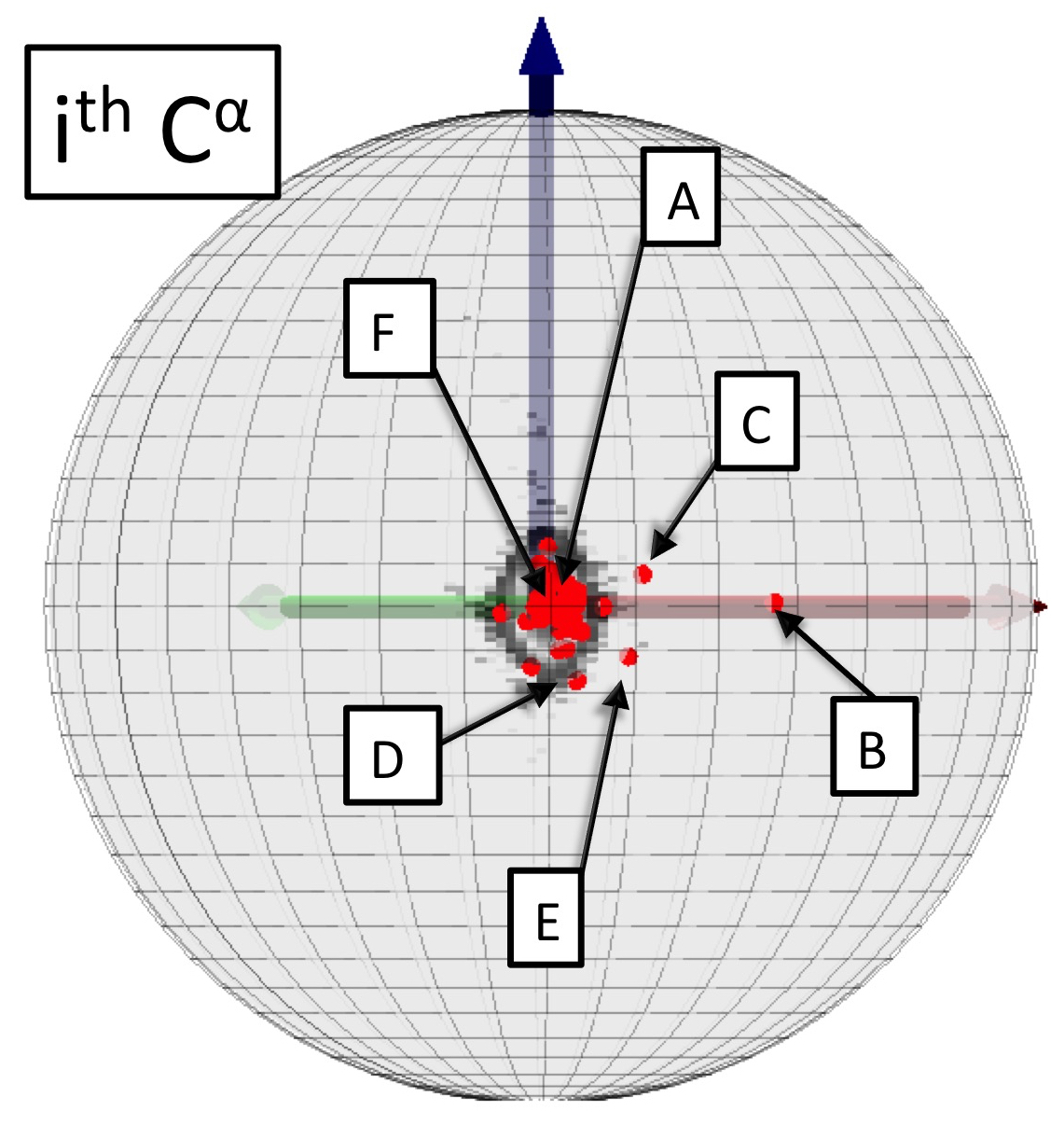}}
                    \quad\quad\quad
          % \subfigure
          {\includegraphics[width=0.4\textwidth]{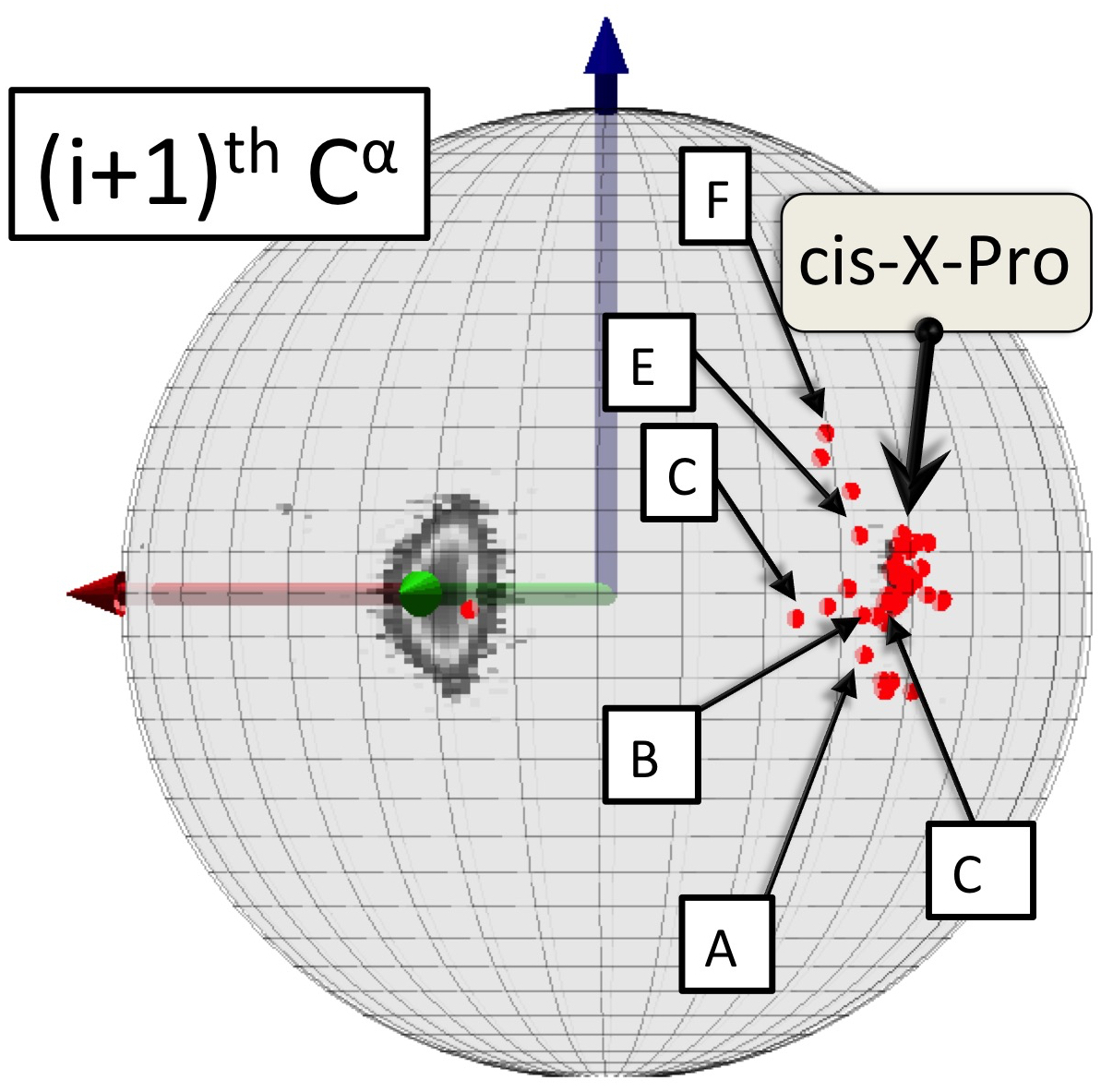}}
         \caption{{ 
      (Color online) Distribution of {\it cis} X--Xnp, for  C$^\alpha_i$ (left) and C$^\alpha_{i+1}$ (right) in C-centered CNO frames, in the background of all PDB C$^\alpha$-structures in our dataset.  }}
        \label{fig11}
 \end{figure}
%%%
%
%
%
%

In Figure \ref{fig11} we show the distributions of the C$^\alpha$ atoms at the vertices of the  {\it cis} X--Xnp peptide plane, in the grey-scaled background of the 1.0 \AA~PDB data, as seen from the position of the C atom in the 
$i^{th}$ CNO frame. The distributions are highly localised and confirm the planar character of the peptide plane.  The distribution of the {\it cis} X--Xnp structures is also centered in this region, but with a much larger dispersion.

\begin{figure}[h]
         \centering
          % \subfigure
          {\includegraphics[width=0.4\textwidth]{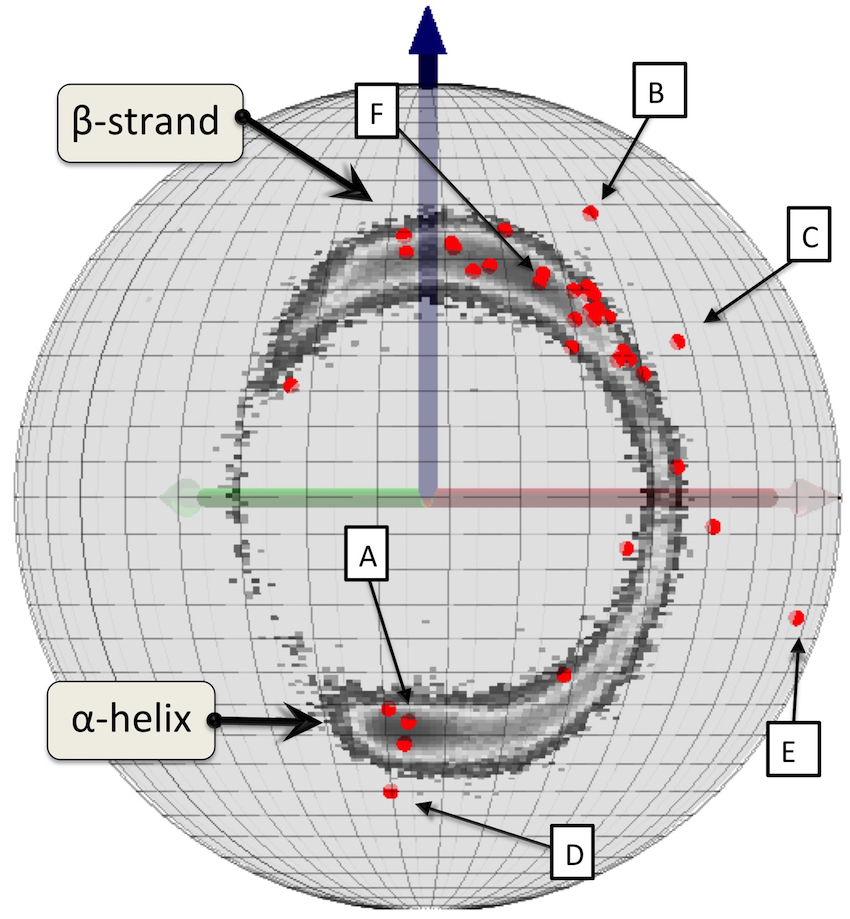}}
          \quad\quad\quad
          % \subfigure
          {\includegraphics[width=0.4\textwidth]{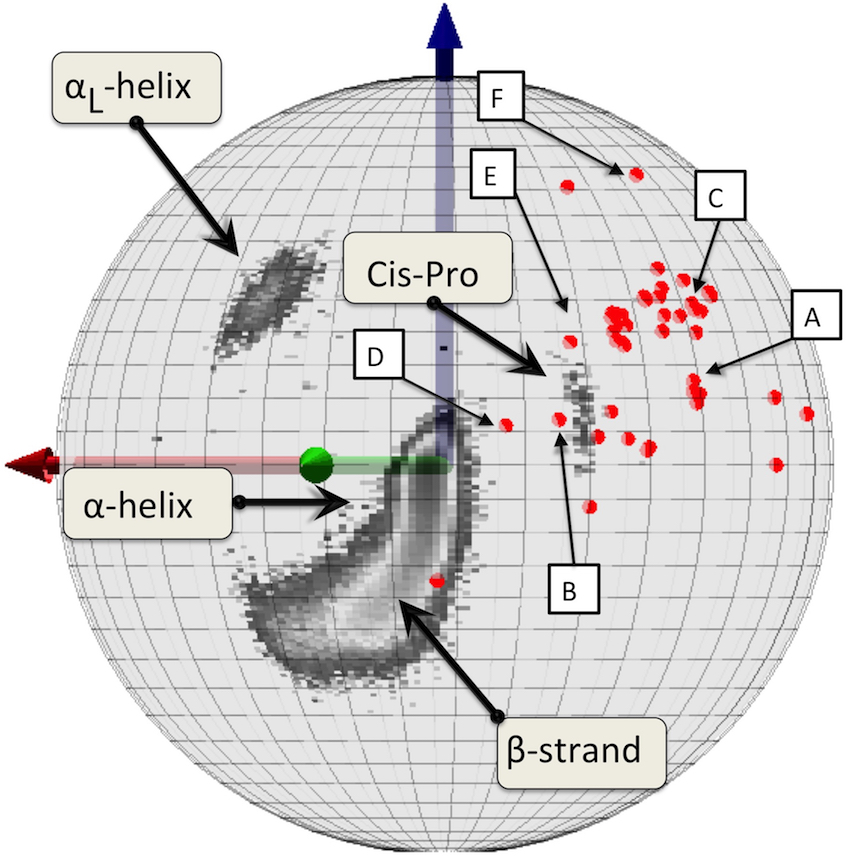}}
         \caption{{ 
      (Color online) Distribution of  C$^\beta_{i}$ (left) and  C$^\beta_{i+1}$ (right) atoms for the 
      $i^{th}$ {\it cis} X--Xnp peptide planes, 
      in the C-centered
      CNO coordinate system.  The grey background is the PDB distribution.  }}
        \label{fig12}
 \end{figure}

In Figure \ref{fig12} we show the distributions of the C$^\beta$ atoms at the vertices of the  {\it cis} X--Xnp peptide plane, in the grey-scaled background of the 1.0 \AA~PDB data, as seen from the position of the C atom in the $i^{th}$ CNO frame. It is notable that in the case of C$^\beta$ we observe a distinct localised {\it cis}--Pro region  in the $i^{th}$ vertex, while in the case of C$^\alpha$  there is such a localised region at the $(i+1)^{th}$ vertex.

%%%%%%%%%%%%%%%%%%%%%%%%%%%%%%%%%%%%%%%%%%%%%
%
%
%
%
%
%figure 9
%%%
%
%
%
%%%%%%%%%%%%%%%%%%%%%%%%%%%%%%%%%%%%%%%%%%%%%

\begin{figure}[h]
         \centering
         % \subfigure
         {\includegraphics[width=0.4\textwidth]{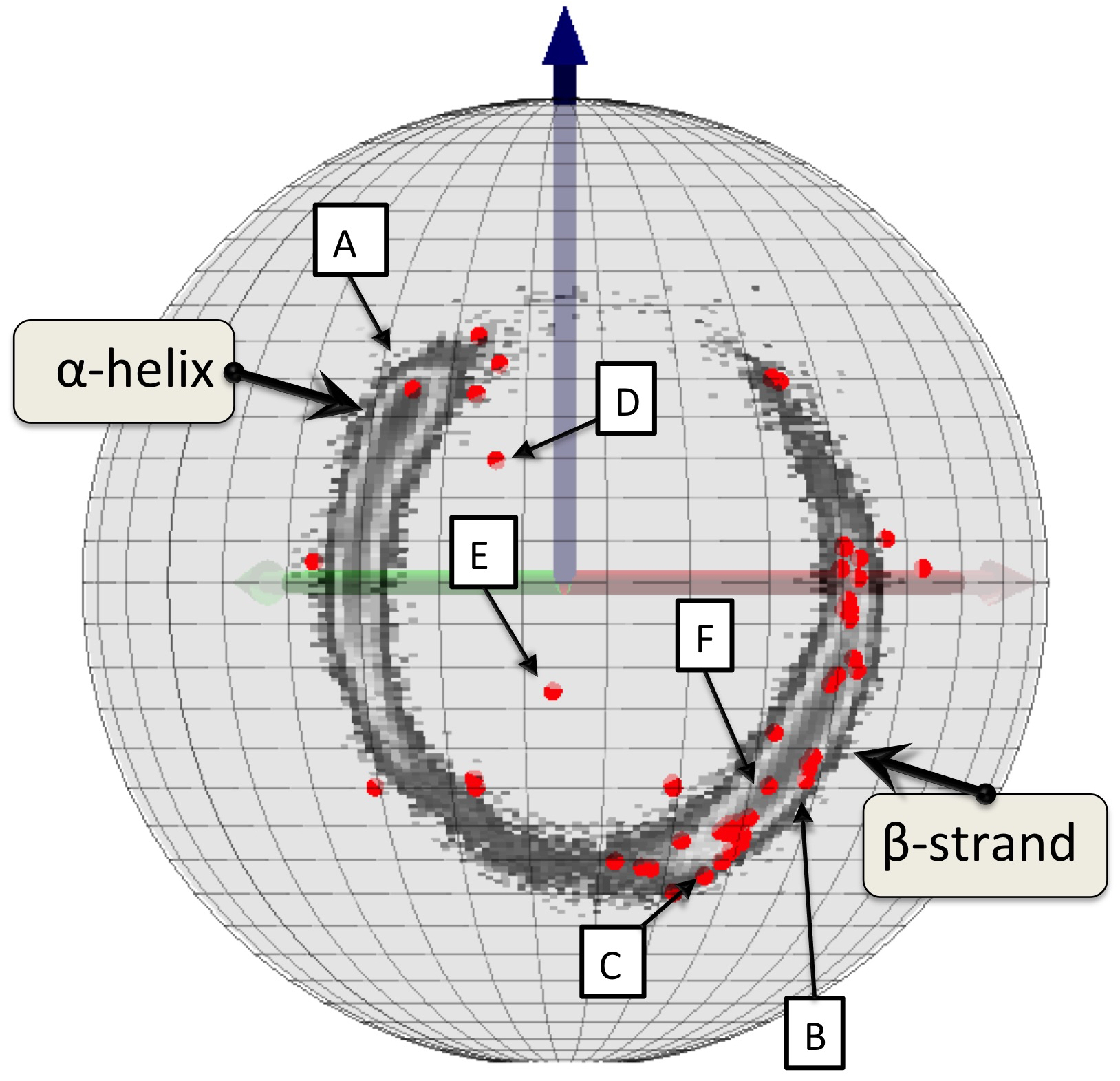}}
         \quad\quad\quad
         % \subfigure
         {\includegraphics[width=0.4\textwidth]{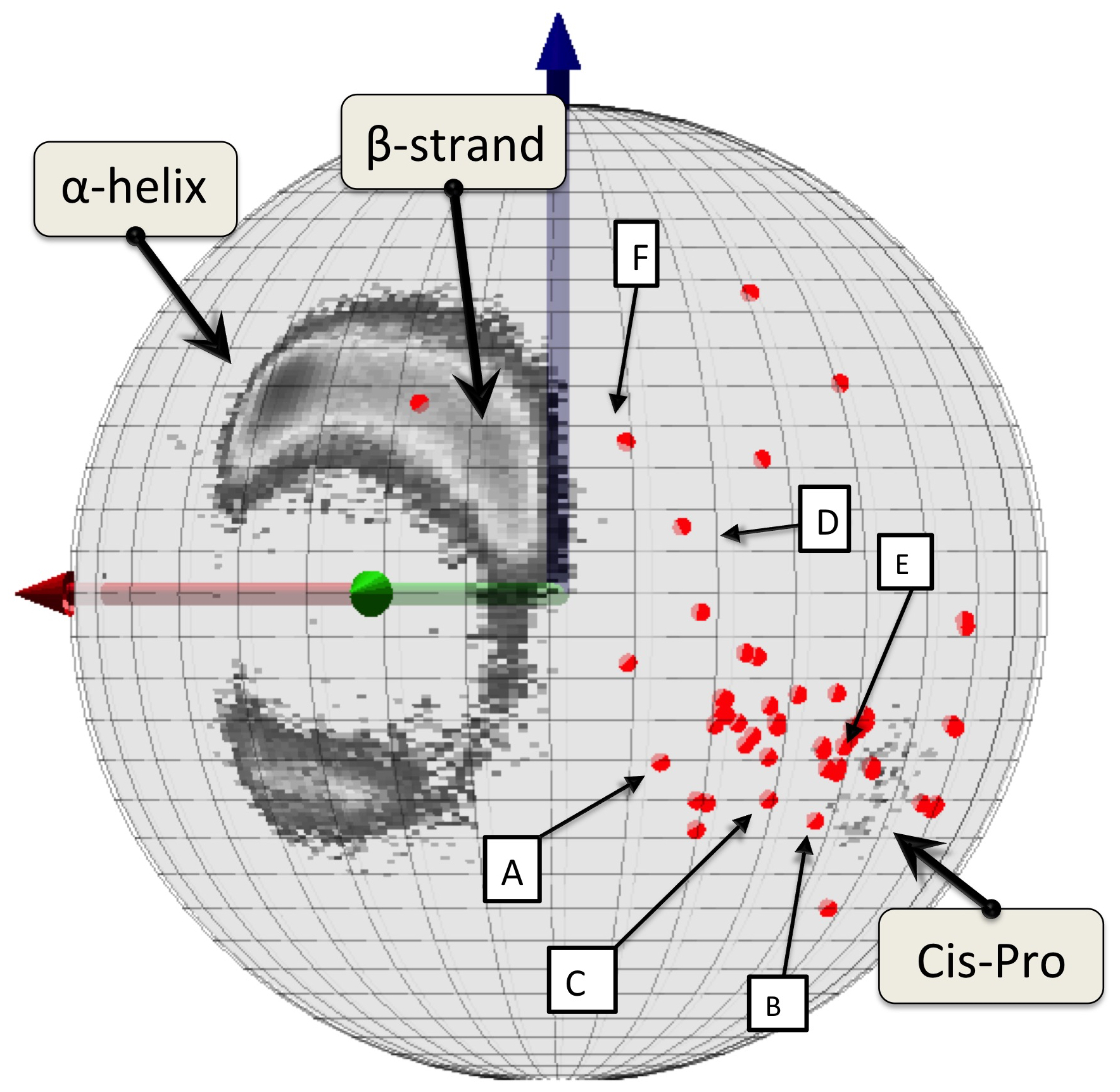}}
         \caption{{ 
      (Color online) Distribution of  N$_{i-1}$ atoms (left) and  C$_{i+1}$ atoms (right) for the $i^{th}$ {\it cis} X--Xnp peptide planes, 
      in the C-centered CNO coordinate system.  The grey background is the PDB distribution.  }}
        \label{fig13}
 \end{figure}

In Figure \ref{fig13} we show the distributions of the N atoms in the $(i-1)^{th}$ peptide plane and the
C atoms in the $(i+1)^{th}$ peptide plane,  as seen from the position of the C atom in the $i^{th}$ CNO frame. We observe that most of the 
{\it cis}-N$_{i-1}$ atoms have positions that are in line with the PDB distribution,  while there is a relatively
large dispersion in the C$_{i+1}$ atoms of the {\it cis} X--Xnp peptide planes that is not observed in the case of {\it cis} X--Pro; we observe  a  distinct localised {\it cis}--Pro region  in the 
ensuing C$_{i+1}$ distribution. 

%%%%%%%%%%%%%%%%%%%%%%%%%%%%%%%%%%%%%%%%%%%%%
%
%
%
%
%
%figure -14_15
%%%
%
%
%
%%%%%%%%%%%%%%%%%%%%%%%%%%%%%%%%%%%%%%%%%%%%%
%
%

By comparing Figures \ref{fig11}--\ref{fig13} we conclude that the $i^{th}$ { \it cis} peptide plane affects
the $i^{th}$ side-chain  (C$^{\beta}_i$) and the $(i+1)^{th}$ peptide plane including C$^{\alpha}_{i+1}$ but has no observable
anomalous effect on C$^{\alpha}_{i}$  and the preceding peptide plane, nor on the $(i+1)^{th}$ side-chain (at 
the level of C$^\beta$). 
%Moreover, unlike the {\it cis} X--Xnp peptide planes, which give rise to a uniform 
%distribution,  the {\it cis} X--Xnp peptide planes display much higher dispersion.
%

\section{Comparison of CNO frames and Ramachandran Angles}

%\subsection*{Comparison with Ramachandran Angles}
 
 A comparison  is due between the (intrinsic)  
angles ($\theta_i,\varphi_i$) of  the CNO frame, and the (extrinsic) Ramachandran angles ($\psi_i, \phi_{i+1}$): 
We recall that $\psi_i$ is a dihedral rotation around the bond  
$\langle$C$^\alpha_i$ -- C$_i$$\rangle$,
while $\phi_{i+1}$ is a dihedral rotation around the bond $\langle$N$_i$ -- C$^\alpha_{i+1}\rangle$. See Figure \ref{fig3}.
Unlike  ($\theta_i,\varphi_i$) which are spherical latitude and longitude, 
the Ramachandran angles are toroidal coordinates and as such 
they do not directly determine the 3D positions of individual atoms: The topology of a torus is different from the topology of a sphere. 
To compare the information content in the Ramachandran angles  to 
that in the spherical CNO coordinates, we evaluate as examples
the ($\theta_i,\varphi_i$) distributions of the N$_{i-1}$ atoms, with the goal  to search for correlations with the Ramachandran angle
 $\psi_i$. In the case of the $\phi_{i+1}$ Ramachandran angle, as an example we then search for correlations with 
 the C$_{i+1}$ coordinate values of ($\theta_i,\varphi_i$). Note that the information content of the C$_{i+1}$ positions in
 terms of $\psi_i$, and that of N$_{i-1}$ in terms of $\phi_{i+1}$ is corrupted by mutual interference.
  
  %%%%%%%%%%%%%%%%%%%%%%%%%%%%%%%%%%%%%%%%%%%%%
  %
  %
  %
  %
  %
  %figure 11
   \begin{figure}[h]
           \centering
           % \subfigure
           {\includegraphics[width=0.4\textwidth]{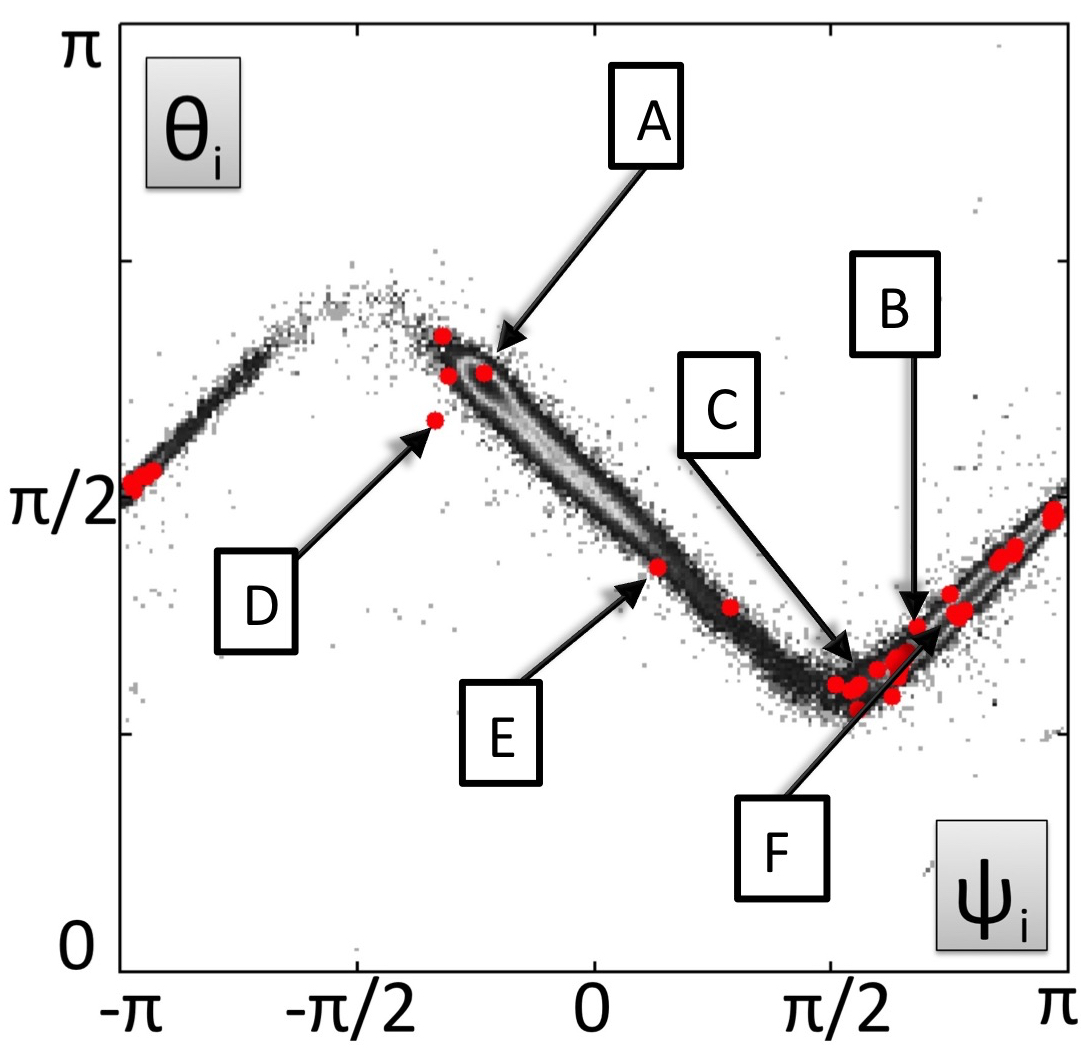}}
           \quad\quad 
           % \subfigure
           {\includegraphics[width=0.4\textwidth]{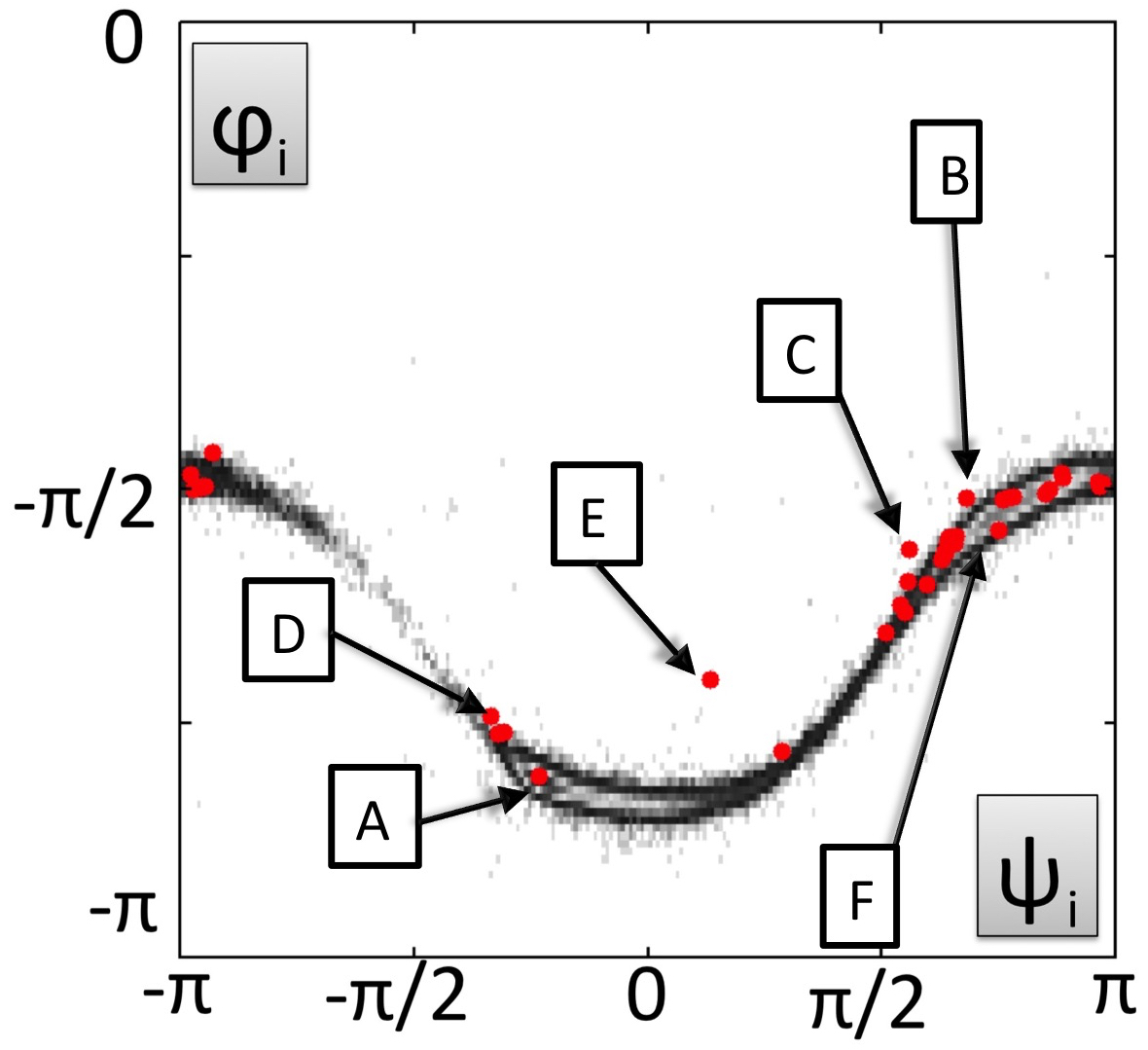}}
           \caption{{ 
        (Color online) (top) Comparison between the Ramachandran angle $\psi_i$, and the latitude angle $\theta_i$ of
        the N$_{i-1}$ atom in the CNO frame two-sphere. (bottom) Comparison between the Ramachandran angle $\psi_i$, 
        and the longitude angle $\varphi_i$ of  the N$_{i-1}$ atom in the CNO frame two-sphere. 
         }}
          \label{fig14}
   \end{figure}
  %%%
  %
  %
  %
  %%%%%%%%%%%%%%%%%%%%%%%%%%%%%%%%%%%%%%%%%%%%%
  % 
  %
 
In Figure \ref{fig14} we show the results for the Ramachandran angle $\psi_i$. We observe a clear correlation
between both CNO coordinates and the Ramachandran angle; since ($\theta_i,\varphi_i$)  are coordinates on 
the two-dimensional surface of the CNO-sphere, their combined information content is then more proliferate than that of the single Ramachandran  angle; the information content in the CNO sphere is visually more detailed.

\begin{figure}[h]
         \centering
          % \subfigure
         {\includegraphics[width=0.4\textwidth]{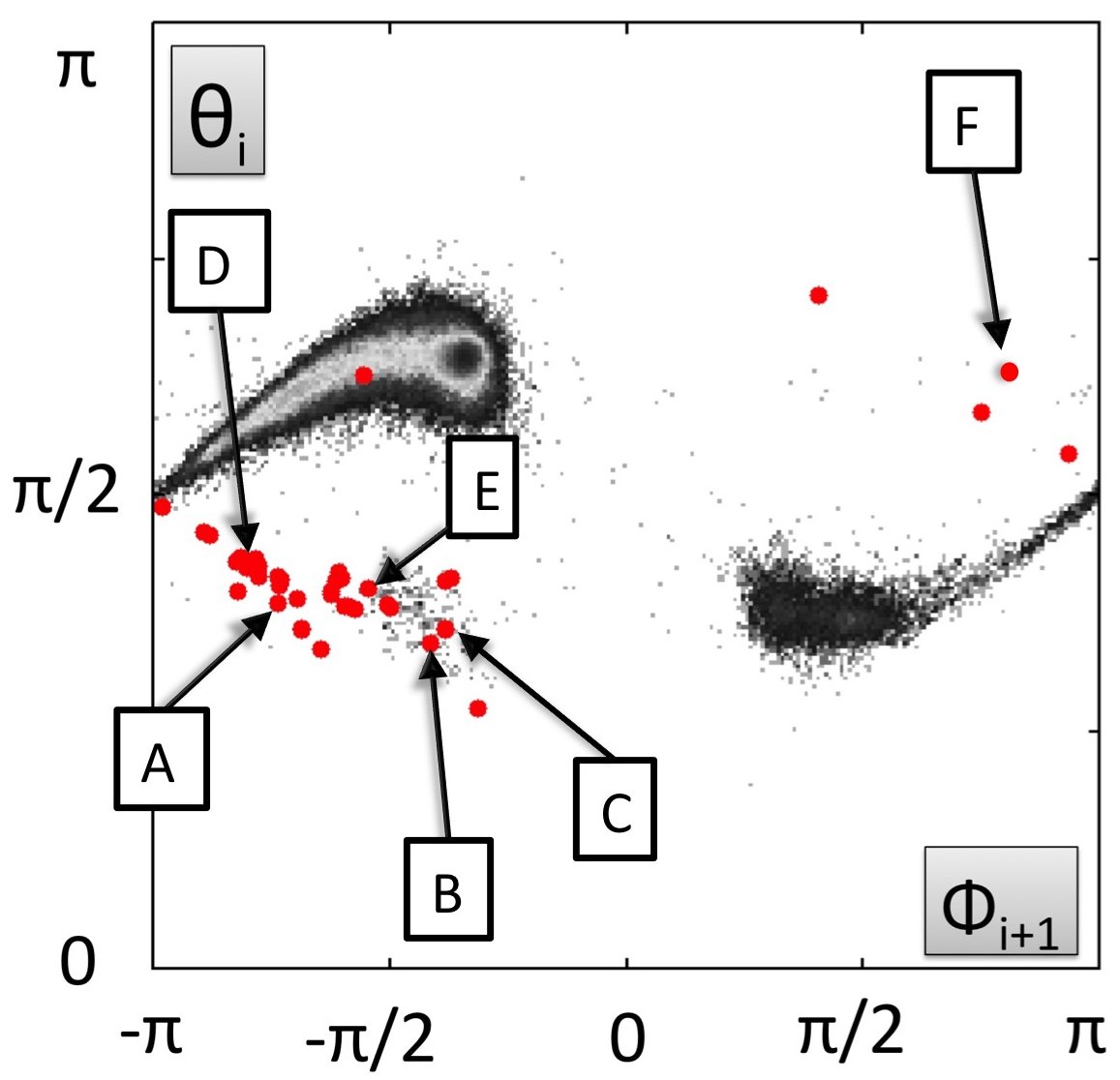}}
         \quad\quad
          % \subfigure
         {\includegraphics[width=0.4\textwidth]{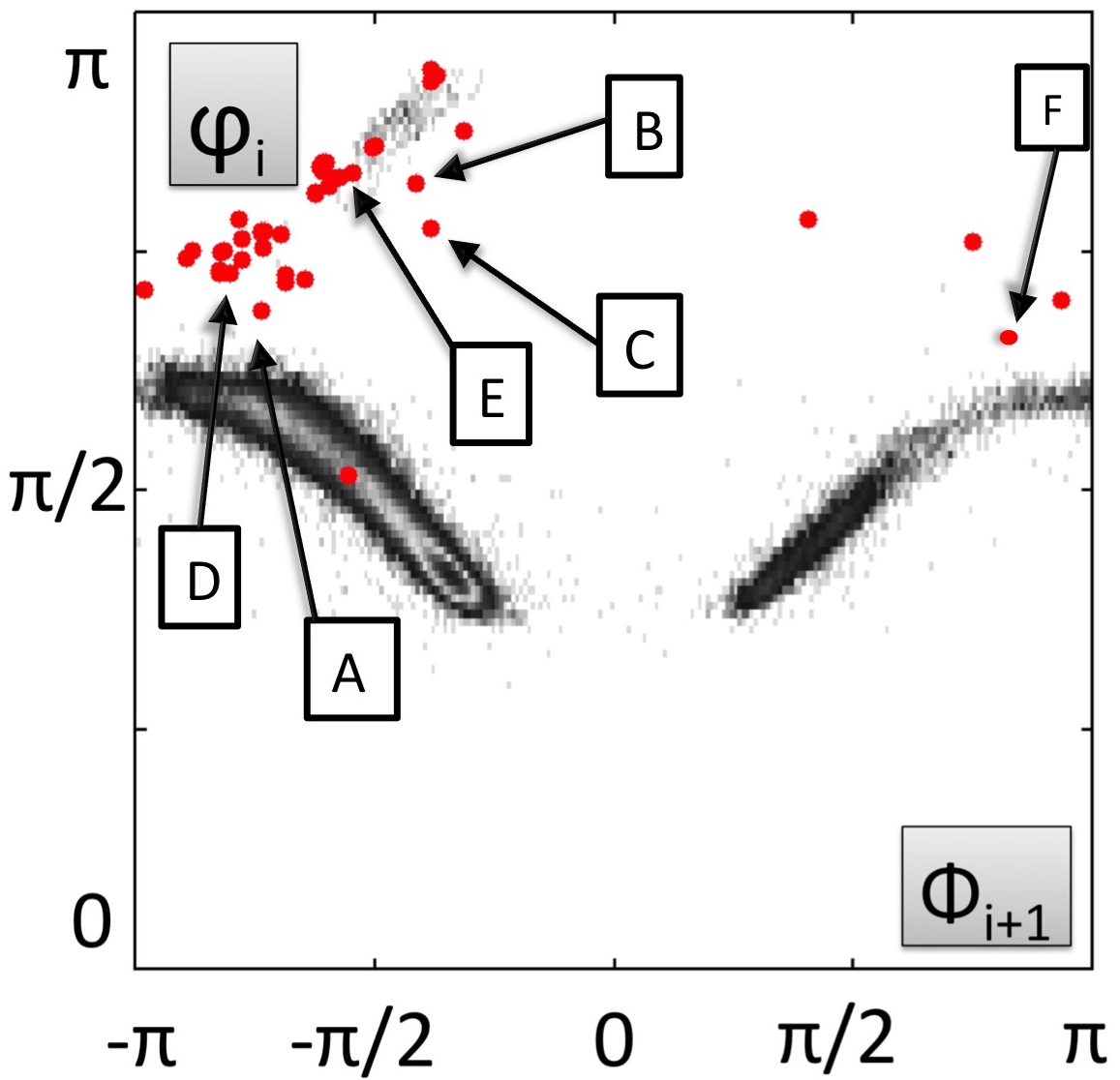}}
         \caption{{ 
      (Color online) (top) Comparison between the Ramachandran angle $\phi_{i+1}$, and the latitude angle $\theta_i$ of
      the C$_{i+1}$ atom in the CNO frame two-sphere. (bottom) Comparison between the Ramachandran angle $\phi_{i+1}$, 
      and the longitude angle $\varphi_i$ of  the C$_{i+1}$ atom in the CNO frame two-sphere. 
       }}
        \label{fig15}
 \end{figure}

In Figure \ref{fig15} we show the results for the Ramachandran angle $\phi_{i+1}$. Also here we observe a strong correlation between both CNO coordinates and the Ramachandran angle. 

%%%%%%%%%%%%%%%%%%%%%%%%%%%%%%%%%%%%%%%%%%%%%
%
%
%
%
%
%figure -14_15
%%%
%
%
%
%%%%%%%%%%%%%%%%%%%%%%%%%%%%%%%%%%%%%%%%%%%%%
%
%

In these figures we have again indicated the individual {\it cis} X--Xnp   entries,  distributed very much in line with the 1.0 \AA~ background.

\section{Summary}
%\subsection*{Conclusions}

We have investigated protein structure both in terms of 
\textit{extrinsic} geometry-based coordinates and \textit{intrinsic}, protein structure based coordinates. As
an example of the former we have considered both Frenet frames and Ramachandran angles, and as an example
of the latter we have introduced and analysed the CNO coordinate system.  We propose to develop  
the CNO coordinates
into a 3D virtual reality based structural analysis method of the protein structure. Indeed, our results suggest
that intrinsic frames are a good complement to  extrinsic frames, they can provide an alternative to the laboratory 
frames that are employed widely in 
3D protein visualisation programs such as VMD, Jmol, PyMOL and others \cite{wiki-viewer}.  
Moreover, a combination of extrinsic and intrinsic coordinates  would yield  versatile
{\it what-you-see-is-what-you-have} type insight 
to the  structural geometry of a protein. In particular, we have demonstrated how such a combination 
can be used to visually detect and analyse anomalously positioned atoms in crystallographic proteins;
as an example we have scrutinised the {\it cis} peptide planes in high resolution PDB structures. 
The methodology we have outlined 
should be of value both in refining low-resolution data and in detecting structural anomalies 
with potentially important biomedical consequences.

\section*{Acknowledgements}
This work was supported in part by Bulgarian Science Fund (Grant DNTS-CN-01/9/2014), Vetenskapsr\aa det (Sweden), Carl Trygger's Stiftelse and Qian Ren Grant at Beijing Institute of Technology.

\end{document}